%% file: ms.tex
\documentclass[manuscript]{aastex}

\usepackage{graphics}
\usepackage{mathptmx}
\usepackage{natbib}
\usepackage[usenames,dvips]{color}

\newcommand{\msun}{$M_\odot$}

\newcommand{\hi}{H\,{\sc i}\rm}

\newcommand{\sii}{Si\,{\sc i}}

\newcommand{\mgi}{Mg\,{\sc i}}

\newcommand{\fei}{Fe\,{\sc i}}
\newcommand{\feii}{Fe\,{\sc ii}}
\newcommand{\tii}{Ti\,{\sc i}}
\newcommand{\tiii}{Ti\,{\sc ii}}

\newcommand{\Ti}[5]{\mbox{$#1\,^#2{\rm #3}^{{\rm #4}}_{\rm #5}$}}
\newcommand{\Fe}[5]{\mbox{$#1\,^#2{\rm #3}^{{\rm #4}}_{\rm #5}$}}
\newcommand{\Teff}{T_{\rm eff}}
\newcommand{\EW}{W_{\lambda}}
\newcommand{\mA}{{\rm m\AA}}
\newcommand{\Elow}{E_{\rm low}}
\newcommand{\Eup}{E_{\rm up}}

\newcommand{\opd}{\log \tau_{\rm 500}}

 \newcommand{\hii}{\ion{H}{2}}

\slugcomment{To appear in Astrophysical Journal}

\shorttitle{Red Supergiants and J-band NLTE effects}
\shortauthors{Bergemann et al.}

\begin{document}


\title{Red Supergiant Stars as Cosmic Abundance Probes:\\
    NLTE Effects in J-band Iron and Titanium Lines}


\author{Maria Bergemann}
\affil{Max-Planck-Institute for Astrophysics, Karl-Schwarzschild-Str.1, D-85741
Garching, Germany}
\email{mbergema@mpa-garching.mpg.de}
\author{Rolf-Peter Kudritzki\altaffilmark{1,2}}
\affil{Institute for Astronomy, University of Hawaii, 2680 Woodlawn Drive,
Honolulu, HI 96822}
\email{kud@ifa.hawaii.edu}
\author{Bertrand Plez}
\affil{Laboratoire Univers et Particules de Montpellier, Universit\'e
Montpellier 2, CNRS, F-34095 Montpellier, France}
\email{bertrand.plez@univ-montp2.fr}
\author{Ben Davies}
\affil{Institute of Astronomy, Univerity of Cambridge, Madingley Road,
Cambridge, CB3 OHA, UK}
\email{bdavies@ast.cam.ac.uk}
\author{Karin Lind}
\affil{Max-Planck-Institute for Astrophysics, Karl-Schwarzschild-Str.1, D-85741
Garching, Germany}
\email{klind@mpa-garching.mpg.de}

\and
\author{Zach Gazak}
\affil{Institute for Astronomy, University of Hawaii, 2680 Woodlawn Drive,
Honolulu, HI 96822}
\email{zgazak@ifa.hawaii.edu}


\altaffiltext{1}{Max-Planck-Institute for Astrophysics,
Karl-Schwarzschild-Str.1, D-85741 Garching, Germany}
\altaffiltext{2}{University Observatory Munich, Scheinerstr. 1, D-81679 Munich,
Germany}


\begin{abstract}
Detailed non-LTE calculations for red supergiant stars are presented to
investigate the influence of NLTE on the formation of atomic iron and titanium
lines in the J-band. With their enormous brightness at J-band red supergiant
stars are ideal probes of cosmic abundances. Recent LTE studies have found that
metallicities accurate to 0.15 dex can be determined from medium resolution
spectroscopy of individual red supergiants in galaxies as distant as 10 Mpc. 
The non-LTE results obtained in this investigation support these findings.
Non-LTE abundance corrections for iron are smaller than 0.05 dex for effective
temperatures between 3400K to 4200K and 0.1 dex at 4400K. For titanium the
non-LTE abundance corrections vary smoothly between -0.4 dex and +0.2 dex as a
function of effective temperature. For both elements, the corrections also
depend on stellar gravity and metallicity. The physical reasons behind the
non-LTE corrections and the consequences for extragalactic J-band abundance
studies are discussed.
\end{abstract}


\keywords{galaxies: abundances --- line: formation --- radiative transfer ---
stars: abundances --- stars: late-type --- supergiants}



\section{Introduction}

One of the most promising ways to constrain the theory of galaxy formation 
and evolution in a dark energy and cold dark matter dominated universe is the
determination of the chemical composition of star forming galaxies in the nearby
and high redshift universe. The observations of the relationship between central
metallicity and galactic mass  \citep{lequeux79, tremonti04, maiolino08} and of
metallicity gradients in spiral galaxies \citep{garnett97, skillman98,
garnett04} are intriguing and a clear challenge of the theory, which uses these
observations to test the theoretical predictions of hierarchical clustering,
galaxy formation, merging, infall, galactic winds and variability of star
formation activity and IMF \citep{prantzos00, naab06, colavitti08, yin09,
sanchez09, delucia04, derossi07, finlator08, brooks07, koeppen07, wiersma09,
dave11a, dave11b}.

So far most of our information about the metal content of star forming
galaxies is obtained from a simplified analysis of \hii~region emission lines,
which use the fluxes of the strongest forbidden lines of (most commonly)
[\ion{O}{2}] and [\ion{O}{3}] relative to H$_{\beta}$. This approach is very
powerful and appealing, since these emission line can be observed in galaxies
through the whole universe from low to high redshift. However, it has two
weaknesses. First, this approach yields basically only the oxygen abundance,
which is then taken, in the absence of any other information, as a place holder
for metallicity. Secondly, the abundances depend heavily on the calibration of
the strong line method  \citep{kewley08} and are therefore subject to systematic
uncertainties of up to $0.8$ dex \citep{kud08, bresolin09}.

An alternative approach avoiding these weaknesses is the spectroscopic
analysis of individual supergiant stars. Here, much progress has been made
through the non-LTE low resolution optical spectroscopy of BA blue supergiants
\citep[see e.g.][for application to spiral galaxies in the Local group]{kud08,
kud10, kud11}. However, this technique is restricted to spectroscopy in the
optical. In consequence, it will not be able to take advantage of the fact that
the next generation of extremely large telescopes such as the GMT, TMT, and the
E-ELT will be diffraction limited telescopes at IR wavelengths allowing for
adaptive optics (AO) supported multi-object spectroscopy.

In this sense, the use of red supergiants (RSGs) as extragalactic abundance
probes has a much larger potential. As direct successors of blue supergiants
they have the same masses (8 to 40 \msun\ at the main sequence) and
the same enormous luminosities (10$^{5}$ to 10$^{6}$ L/L${\odot}$), but their
SEDs peak in the infrared, where the gain in limiting magnitude through AO goes
with the fourth power of telescope diameter. Recently, \citet{davies10}
(hereinafter DKF10) have introduced a novel technique, which uses medium
resolution (R $\approx$ 3000) spectroscopy in the J-band to derive stellar
parameters of RSGs with the accuracy of $\sim$ 0.15 dex per individual star.
The J-band spectra of of RSGs are dominated by strong and isolated atomic lines
of iron, titanium, silicon and magnesium, while the molecular lines of OH,
H$_{2}$O, CN, and CO which plague the H- and K-bands are weak and appear as a
pseudo-continuum at R $\sim 3000$ resolution. The prospects of the J-band
technique in application to RSGs were discussed by \citet{evans11}, who showed
that with future instruments it would be possible to measure abundances of
various chemical elements out to enormous distances of 70 Mpc.

In the view of these perspectives, it is important to investigate the
DKF10 method in more detail. One crucial question is how model atmosphere and
line formation calculations are affected by the departures from Local
Thermodynamic Equilibrium (LTE), which are expected because of extremely low
gravities and hence low densities of RSG atmospheres. Detailed non-LTE line
formation calculations for RSGs focussing on the key diagnostic lines in the
J-band are, thus, a first step. In this paper, we present the first calculations
of this type and investigate the formation of iron and titanium lines and the
influence of non-LTE effects on the determination of abundances in the J-band.
In future work, we plan to extend these studies to the formation of silicon and
magnesium lines, which are important for the determination of effective
temperature in the DKF10 method.

The paper is structured as follows. In sections 2 and 3 we describe the model
atmospheres and the details of the line formation calculations. Section 4
presents the results: departure coefficients, line profiles and equivalent
widths in LTE and non-LTE and non-LTE abundance corrections for chemical
abundance studies. Section 5 discusses the consequences for the new J-band
diagnostic technique and aspects of future work.

\section{Model Atmospheres}

The atmospheric structure for our non-LTE line formation calculations is
provided by the MARCS model atmospheres. The physical assumptions
underlying these atmospheres are described in \citet{gustafsson08}. In short,
these LTE models are spherically-symmetric, in 1D hydrostatic equilibrium and
include convection within the framework of the mixing-length theory. A careful
discussion of the strengths and the weaknesses of these models is given by
\citet{gustafsson08, plez10}. The reference solar abundance mixture in these
models is that of \citet{grevesse07}.

For our investigation we use a small grid of models computed assuming the mass
of 15 \msun with five effective temperatures (T$_{\rm eff}$ = 3400, 3800,
4000, 4200, 4400K), three gravities ($\log g = 1.0$, $0.0$, $-0.5$ (cgs)), three
metallicities ([Z]$\,\equiv\,$ log Z/Z$_{\odot}$ $= -0.5$, $0.0$, $+0.5$). We
also use two values of micro-turbulence $\xi_{t} = 2$ and 5 km/s, respectively.
This grid covers the range of atmospheric parameters expected for RSGs (see
DKF10). We also compared the 1 and 15 \msun models for the same parameters
finding very small differences in the NLTE abundance corrections.

For the current analysis, the model atmospheres were carefully extrapolated
to a continuum optical depth at $500$ nm $\opd = -5$, since for certain
combinations of RSG atmospheric parameters the cores of the Ti and Fe lines have
contributions from atmospheric layers around optical depth $\opd \sim -3.5$,
which is close to the top boundary of the MARCS atmospheric models ($\opd =
-4.0$). By this extrapolation we explored whether higher layers could contribute
to the formation of the IR lines investigated. No significant effects were
found.

At the highest effective temperature (T$_{\rm eff}$ = 4400K) and 
lowest gravity ($\log g = -0.5$) the MARCS models show a very slight temperature
inversion, although they are converged. The locations of these inversions can be
noticed from the kinks in the plots of departure coefficients (Fig. \ref{dep1}
and \ref{dep2}, see Sect. 3.1 and 3.2) for \fei and \tii. They do not have any
significant effects on the results.

\section{Non-LTE Line Formation}

\subsection{Statistical equilibrium calculations}

For the non-LTE line formation calculations we use DETAIL, a non-LTE
code widely used for hot stars \citep{przybilla06} as well as for cool stars
\citep{butler85}. DETAIL is a well established and well tested code, which is
fast and very efficient through the use of the accelerated lambda iteration
scheme in the formulation of \citet{1991A&A...245..171R,1992A&A...262..209R},
and allows for the self-consistent treatment of overlapping transitions and
continua. It is worth noting that DETAIL requires the partial pressures of all
atoms and important molecules to be supplied with a model atmosphere; these were
computed using the MARCS equation-of-state package.

\begin{figure*}[ht!]
\begin{center}
\resizebox{0.8\textwidth}{!}{\includegraphics[scale=1]
{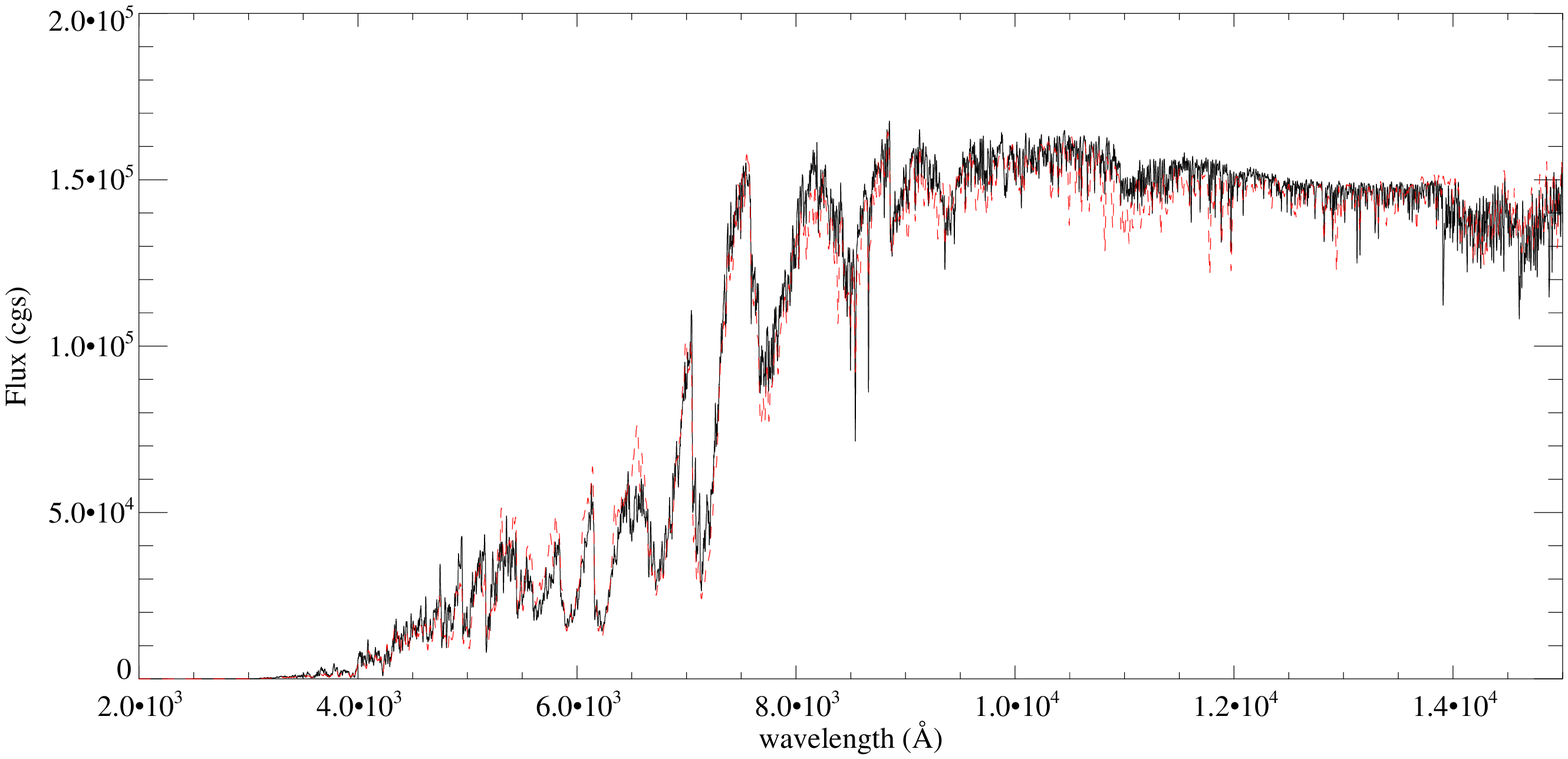}}
\resizebox{0.8\textwidth}{!}{\includegraphics[scale=1]
{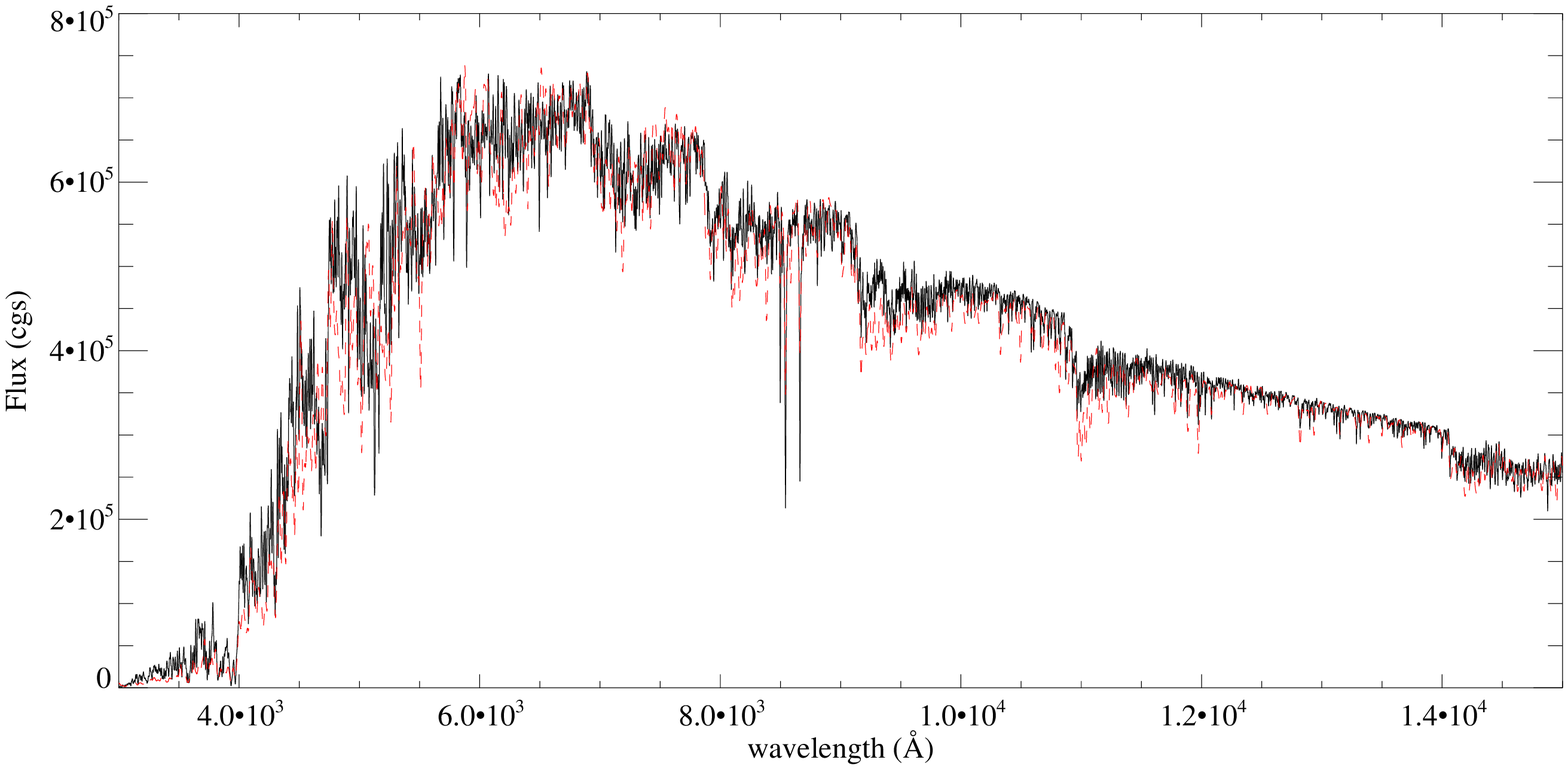}}
\caption{Emergent fluxes computed with DETAIL (black solid) as compared with the
original SED's from MARCS (red, dashed). Top: $\Teff = 3400$ K, $\log g = 0$,
[Z] $ = 0$. Bottom: $\Teff = 4400$ K, $\log g = 0$, [Z] $ = 0$.}
\label{sed}
\end{center}
\end{figure*}

Since the non-LTE effects depend crucially on the radiation field at all
wavelengths, it is important that all relevant opacities are included in the
statistical equilibrium calculations. The background line opacity is computed
directly for each wavelength in the model atom using extensive linelists
extracted from the NIST \citep{nist} and
Kurucz\footnote{http://kurucz.harvard.edu} databases. For the current project,
these tables were complemented by the TiO line data of
\citet{1998A&A...337..495P} to account for the extreme line blanketing by TiO
molecules. A comparison of spectral energy distributions calculated with DETAIL
and MARCS for two typical RSG models is given in Fig. \ref{sed}. The agreement
is satisfactory.

The non-LTE radiative transfer in DETAIL is done in plane-parallel geometry.
This might be a reason of concern, since the atmospheres of RSGs have a large
scale height because of small $\log g$. \citet{heiter06}
have used the spherically extended MARCS models to compare LTE line formation
calculations with radiative transfer in spherically extended geometry
with the calculations done with the plane-parallel approximation. At
most extreme cases abundance corrections were smaller than $0.1$ dex
indicating that the combination of spherically extended models with plane
parallel radiative transfer yields reasonable results. We note that the lowest
gravities used by \citet{heiter06} were $\log g = 0.5$ (cgs), whereas the lowest
gravities in our grid are a factor of ten lower, i.e. $\log g = -0.5$ (see
below). On the other hand, \citet{heiter06} investigated the most extreme
models with very low stellar mass, i.e. 1 \msun. However, since the spherical
extend of a line forming atmosphere $\Delta$R in units of the photospheric
radius R is:

\begin{equation}
\Delta R /R \sim T_{\rm eff}/(g~M)^{1/2}
\end {equation}

and RSGs are massive stars with masses equal and larger than 10 \msun, we expect
the effects of sphericity to be small. Indeed, compared with Heiter \&
Eriksson's largest extension of 12\%, $\Delta$R/R of our most extreme model
(calculated with 15 \msun) is 7\%. We note however that most likely real RSG
atmospheres are more extended due to hydrodynamical phenomena \citep[][and
references therein]{2011A&A...535A..22C}. We have also checked $\Delta$R/R for
the line forming region of the IR lines of our investigation and found values
less then $2$ \%. This, in addition to a relatively small geometrical dilution
factor of the radiation field ($(1/1.07)^2$ or 14\% in the most extreme model
with $\Delta$R/R = 7\%), leads us to conclude that the NLTE abundance
corrections are likely very similar in plane-parallel and spherical geometry.

The line profiles and the NLTE abundance corrections were computed with a
separate code SIU \citep{reetz} using the level departure coefficients from
DETAIL. SIU and DETAIL share the same physics and line lists.

\subsection{Model atoms}

\begin{figure*}[ht!]
\begin{center}
\includegraphics[scale=0.8, angle=90]{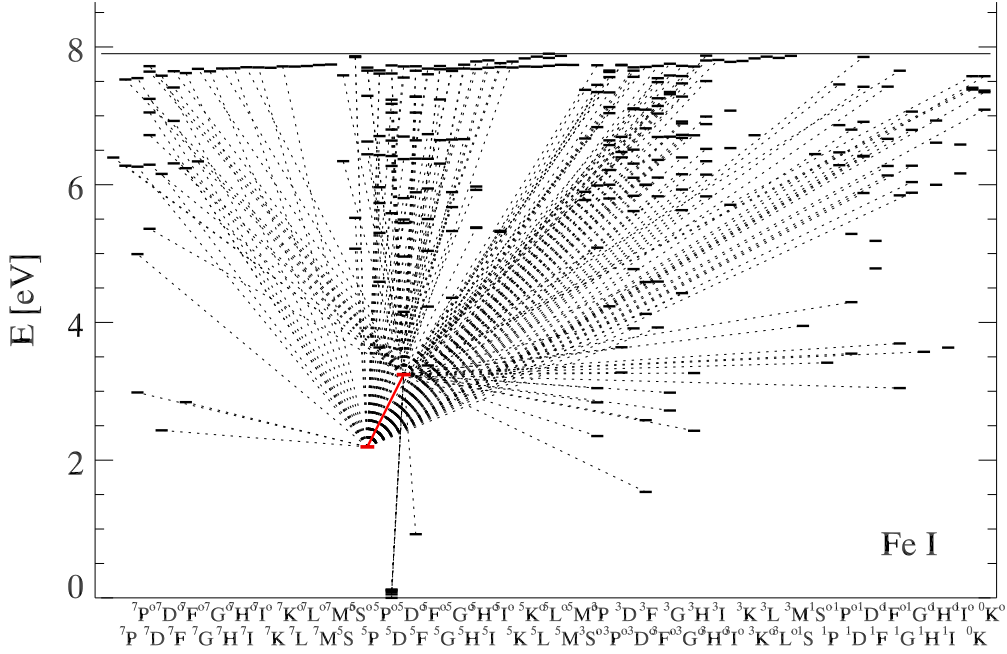}
\caption{The \fei\ non-LTE atomic model. The terms leading to the J-band IR
transitions discussed are highlighted in red. Only radiative transitions to and
from these levels are shown in this plot. The calculations use a much more
complex model of \fei\ with many more transitions and simultaneously also a
detailed atomic model of \feii\, which is also not shown here.}
\label{felev}
\end{center}
\end{figure*}

The atomic models for Ti and Fe are essentially those described in
\citet{bergemann11} and \citet{bergemann12}, however some modifications were
implemented for the current analysis. The models include the first two
ionization stages of each element accounting for $332$ and $397$ levels of Ti
and Fe, respectively. For \fei\, we have added the fine structure splitting of
the \Fe{a}{5}{P}{}{} and \Fe{z}{5}{D}{\circ}{} states, since the J-band IR lines
of our investigation form between these levels. We also removed a number of
theoretically-predicted \fei\ levels and radiative transitions, which according
to our tests are not important for the statistical equilibrium of the element in
RSG atmospheres. The Fe level diagram with transition related to the J-band
lines investigated here is shown in Fig. \ref{felev}. Similarly, the \tii\ model
was extended by the fine structure of all levels below $2.73$ eV, and all
transitions involving these levels were updated (Fig. \ref{felev}). The number
of radiatively-allowed transitions is, thus, $4943$ and $5328$ for Ti and Fe,
respectively; the gf-values of these transitions were taken from various
sources \citep{bergemann11,bergemann12}. Photoionization cross-sections
for \fei\ were kindly provided by  Bautista (2011, private communication); they
are computed on a more accurate energy mesh and provide better resolution of
photoionization resonances compared to the older data, e.g. provided in the
TOPbase. For Ti the hydrogenic approximation was adopted.

To compute collisional rates, we relied on standard recipes, commonly
used in NLTE calculations for late-type stars, as no better alternatives are
available. Electron impact excitation and ionization cross-sections were
computed using the semi-empirical formulae from \citet{1962amp..conf..375S},
\citet{1962ApJ...136..906V}, and \citet{1973asqu.book.....A}. Inelastic
collisions with \hi\ atoms were computed according to the Drawin's
formula in the formulation of \citet{1984A&A...130..319S}. From detailed
comparison with more accurate quantum-mechanical and experimental data for other
atoms, the recipes for e$^-$ collisions are expected to give an order of
magnitude accuracy \citep{mash96, lind11}, whereas the situation with \hi\ 
collisions is more uncertain. Recent quantum-mechanical calculations
suggest that for bound-bound transitions the latter are exaggerated
\citep{barklem11, barklem12}, whereas charge transfer processes can not be
described by the Drawin's formulae at all. We note, however, that only light
atoms, Li, O, Na, and Mg, have been investigated so far.

Here, we performed a set of test calculations varying the collision
cross-sections by several orders of magnitude. As expected, the most influential
factor is inelastic \hi\ collisions, which is also well-known from the
NLTE analysis of late-type stars (Gehren et al. 2001, Bergemann et a. 2012).
However, even a factor of ten variation of these cross-sections leads to a
change in the NLTE corrections for the IR \tii\ lines by less than $0.07$ dex
across the RSG parameter space. A notable effect on the coolest models (3400 -
3800 K) is seen when the bound-bound \hi\ collision rates are decreased by four
orders of magnitude. In this case, the NLTE corrections double. For the \fei\ IR
lines, the effect of varying the \hi\ collisions is very small, on
average of the order $0.03$ dex. In our recent work on low-mass FGK stars
\citep{bergemann12}, we constrained the efficiency of inelastic \hi\ collisions
empirically, from the analysis of stars with parameters determined by
independent methods. We can not follow the same approach in this work, as the
sample of nearby RSG's is very limited and stellar parameters are not accurate
enough to permit calibration of collision efficiency using high-resolution
spectra. A more accurate and realistic quantification of the uncertainties
related to the collision rates will therefore have to await for detailed
theoretical calculations and high-quality observations of RSG's.
\begin{figure*}[ht!]
\begin{center}
\includegraphics[scale=0.8, angle=90]{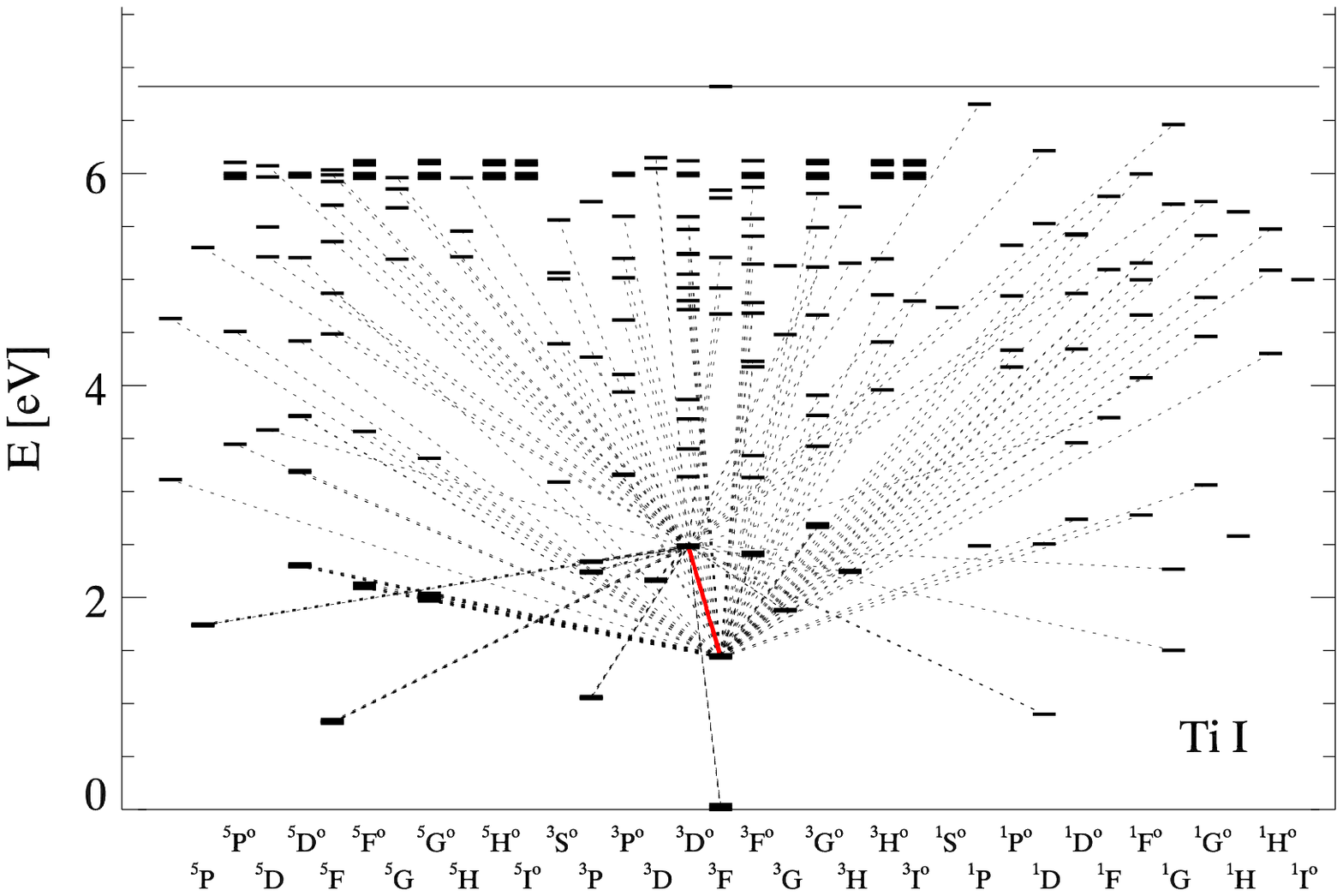}
\caption{The \tii\ non-LTE atomic model. The terms leading to the J-band IR 
transitions discussed are highlighted in red. Only radiative transitions to and
from these levels are shown in this plot. The calculations use a much more
complex model of \tii\ with many more transitions and simultaneously also a
detailed atomic model of \tiii, which is not shown here.}
\label{tilev}
\end{center}
\end{figure*}

\begin{figure*}[ht!]
\begin{center}
\includegraphics[scale=0.3]{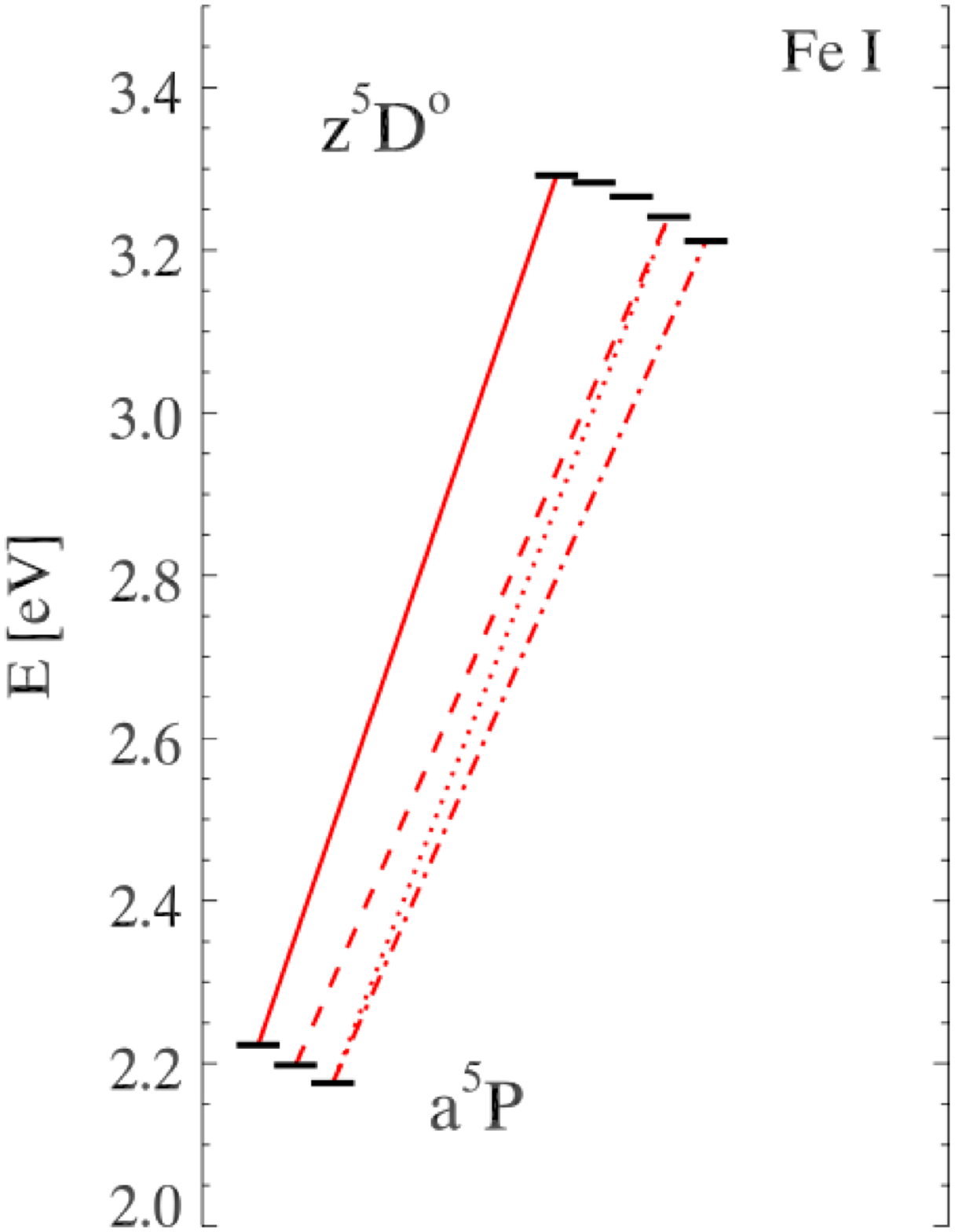}
\includegraphics[scale=0.3]{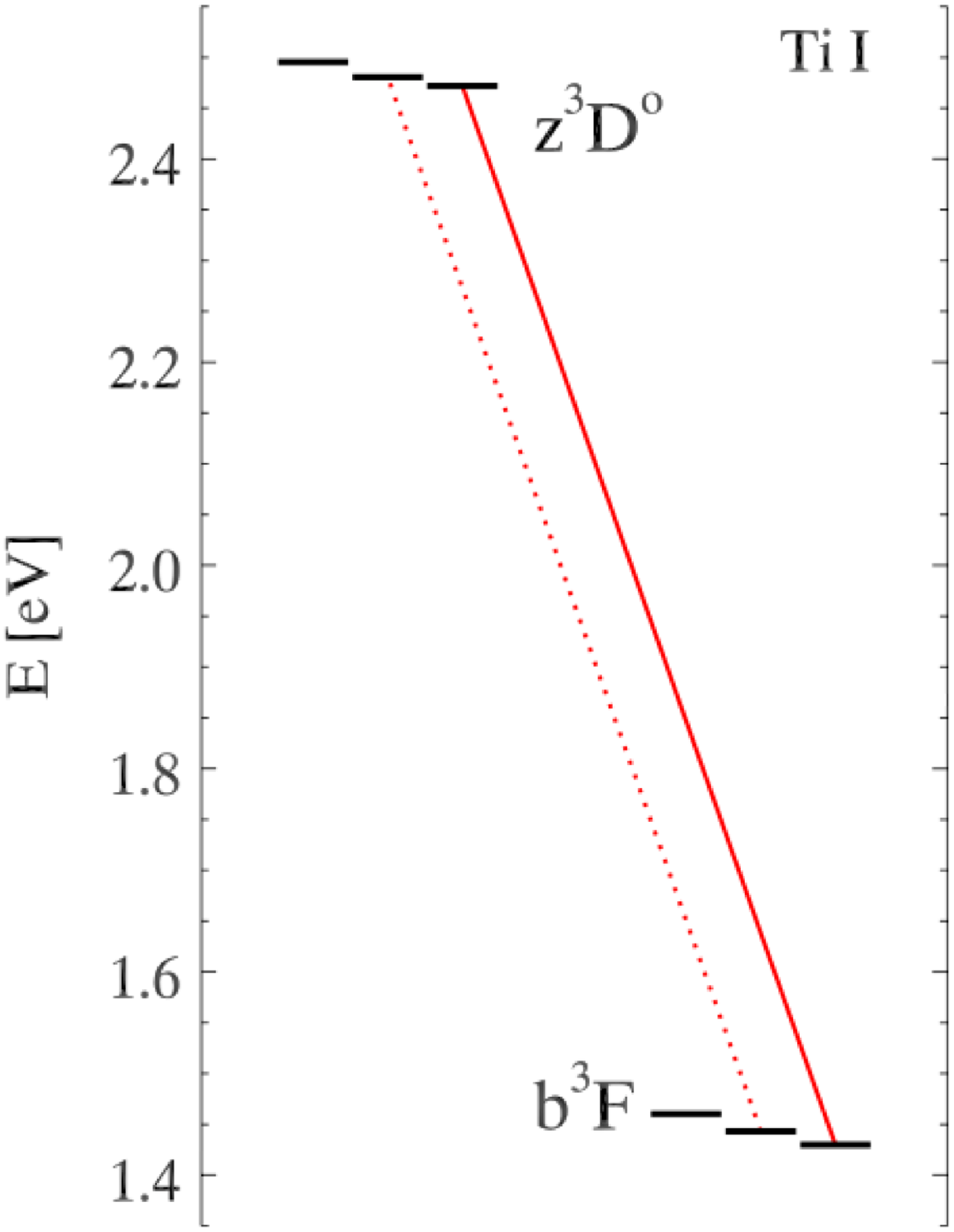}
\caption{Fine structure splitting of \fei\ (left) and \tii\ (right) IR J-band
lines.}
\label{fstruc}
\end{center}
\end{figure*}

\section{Results}

The NLTE effects will be discussed using the concept of level departure
coefficients $b_i$

\begin{equation}
b_i = n_i^{\rm NLTE}/n_i^{\rm LTE}
\end {equation}

where $n_i^{\rm NLTE}$ and $n_i^{\rm LTE}$ are NLTE and LTE atomic level
populations [cm$^{-3}$], respectively.

Furthermore, the importance of the NLTE effects for the determination of
element abundances can be assessed by introducing NLTE abundance corrections
$\Delta_{\rm Fe, Ti}$, where:

\begin{equation}
\Delta_{\rm Fe, Ti} = \log \rm{A (Fe, Ti)}_{\rm NLTE} - \log \rm{A (Fe,
Ti)}_{\rm LTE}
\end {equation}

is the the logarithmic correction, which has to be applied to an LTE iron or
titanium abundance determination A of a specific line to obtain the correct
value corresponding to the use of NLTE line formation. We calculate these
corrections at each point of our model grid for each line by matching the NLTE
equivalent width through varying the Fe abundance in the LTE calculations.
Note that from the definition of $\Delta_{\rm Fe, Ti}$ a NLTE abundance
correction is positive, when for the same element abundance the NLTE line
equivalent width is smaller than the LTE one, because it requires a lower LTE
abundance to fit the NLTE equivalent width.

It shall be kept in mind that both \tii\ and \fei\ have a very complex atomic
structure. For each of the energy levels, there are about a thousand of
radiative and collisional processes, which contribute to the net population or
de-population of a level. As a result, there is a strong interlocking of
radiation field in different lines and continua, and it becomes highly
non-trivial to isolate the processes explaining the populations of individual
atomic levels once the statistical equilibrium has been established.

Thus, in the next two subsections (4.1 and 4.2), we will give only a qualitative
and rather general description of the processes driving departures from LTE in
the excitation-ionization equilibria of Ti and Fe, focusing mainly on the
levels and transitions used in the spectroscopic J-band analysis of the RSG
stars. The selected lines are given in the Table 1. Their wavelengths,
excitation energies, and transition probabilities were extracted from the VALD
database \citep{1995A&AS..112..525P,1997BaltA...6..244R,1999A&AS..138..119K,
2000BaltA...9..590K}. For the \fei\ lines, the transition probabilities are
taken from \citet{1991JOSAB...8.1185O}. Under typical physical conditions in
the RSG models, all these lines are relatively strong, with equivalent widths
$\EW$ exceeding few hundred \mA.

\subsection{Statistical equilibrium of Fe}{\label{sec:fe}} 
\begin{figure*}[ht!]
\begin{center}
\resizebox{0.75\textwidth}{!}{\includegraphics[scale=0.8]
{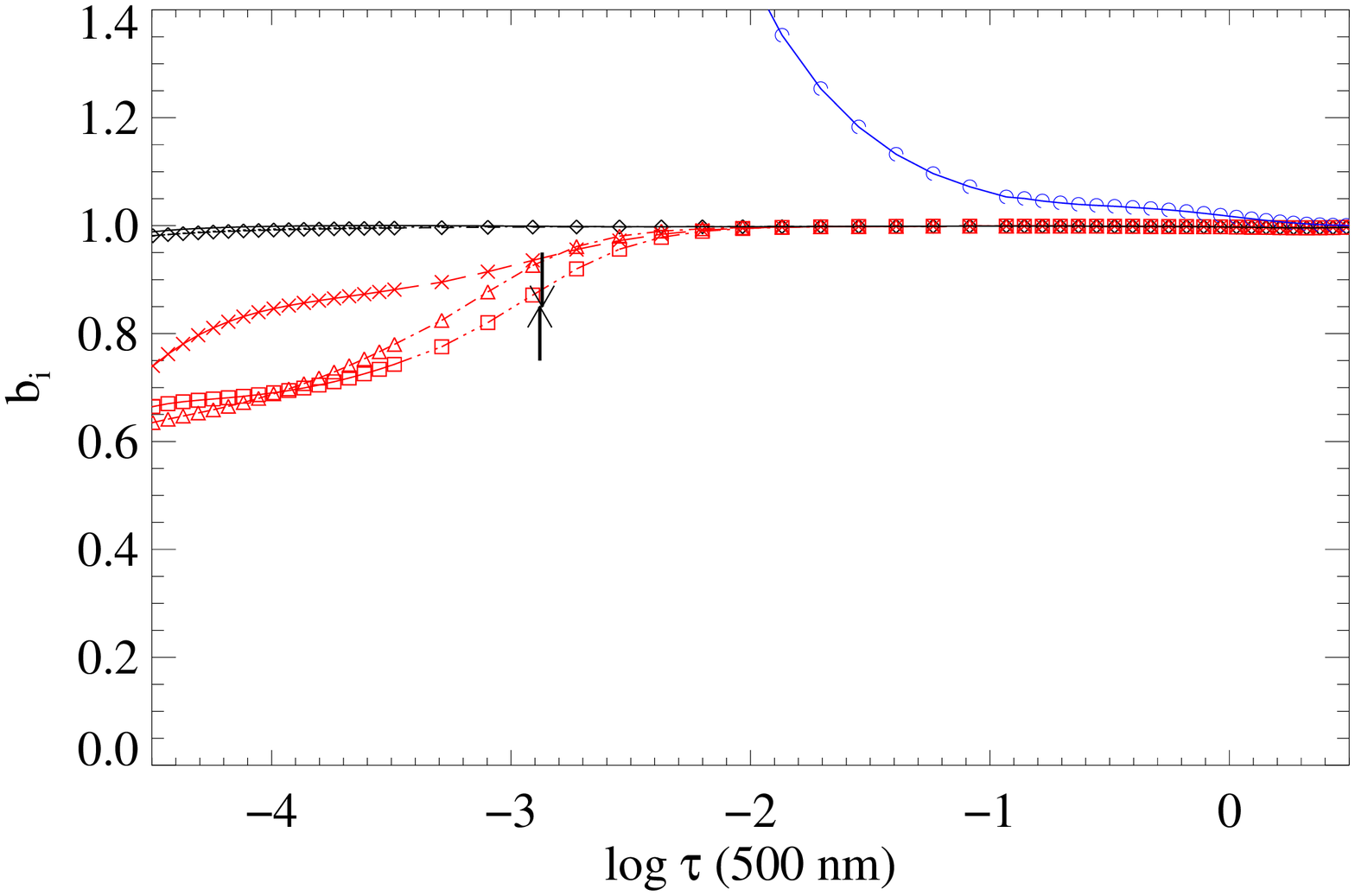}}
\resizebox{0.75\textwidth}{!}{\includegraphics[scale=0.8]
{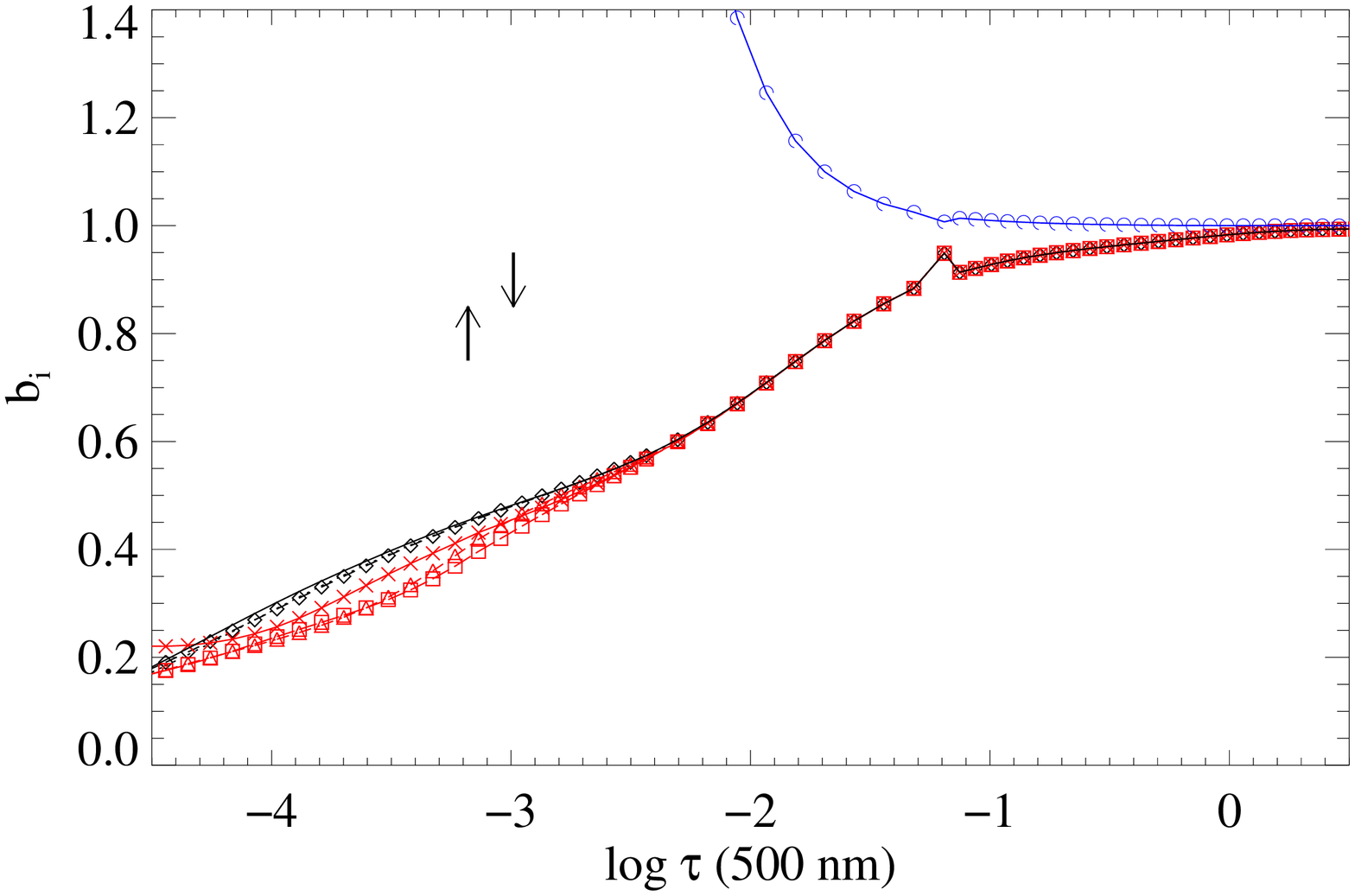}}
\caption{The NLTE departure coefficients of \fei\ for RSG models with log g =
-0.5, [Z] = -0.5 and T$_{\rm eff}$ = 3400K (top) and 4400K (bottom) as a
function of optical depth. Blue:
\feii\ ground state \Fe{a}{6}D{}{9/2}. Black: lower fine structure levels of
J-band IR transitions, \Fe{a}{5}P{}{1} (diamonds, dashed), \Fe{a}{5}P{}{2}
(dotted), \Fe{a}{5}P{}{3} (solid). Red: upper fine structure levels of
IR-transitions, \Fe{z}{5}D{\circ}{0} (crosses), \Fe{z}{5}D{\circ}{3} (squares,
dashed), \Fe{z}{5}D{\circ}{4} (triangles). The LTE and NLTE line core optical
depths $\log \tau (11973$ \AA, \fei$) = 0$ are also indicated by the upward and
downward directed arrows, respectively.}
\label{dep1}
\end{center}
\end{figure*}
Statistical equilibrium of Fe in stellar atmospheres has been extensively
studied in application to solar-type stars, i.e., FGK spectral types
(\citealt{2001A&A...380..645G},\citealt{2011A&A...528A..87M},\citealt{
bergemann12},\citealt{lind12}). However, all theses studies focused on the NLTE
line formation of \fei\ and \feii\ in the near-UV and optical spectral windows.
Here we model the four near-IR \fei\ transitions in the multiplet $296$ (Table
1), which connect the lowest metastable $2.2$ eV \Fe{a}{5}{P}{}{} levels to the
$3.2$ eV \Fe{z}{5}{D}{\circ}{} levels. The departure coefficients for these
selected \fei\ levels are shown in Fig. \ref{dep1} for two selected RSG models.
The LTE and NLTE unity optical depths in the transition at $11973$ \AA\
(\Fe{a}{5}{P}{}{3} $\leftrightarrow$ \Fe{z}{5}{D}{\circ}{4}) are indicated by
the upward and downward directed arrows, respectively.

In the atmospheres of the solar-type stars, the main NLTE effect on the \fei\
number densities is over-ionization by super-thermal UV radiation field, which
reduces the \fei\ number densities compared to LTE. This effect is equally
important for other minority atoms, such as \tii\, which constitute only a tiny
fraction of the total element abundance. For example, everywhere in the solar
photosphere, the ratio of total number densities of neutral to ionized iron, 
$N_{\rm Fe I}/N_{\rm Fe II}$, is always below $0.1$.

The situation is somewhat different in the cool atmospheres of RSGs.
At T$_{\rm eff}$ = 3400K the ratio $N_{\rm Fe I}/N_{\rm Fe II}$ is about
10$^{2}$ at optical depths $\opd \leq -0.5$. In consequence, photoionization
becomes utterly unimportant for the statistical equilibrium of neutral iron at
this low temperature. For T$_{\rm eff}$ = 4400K $N_{\rm Fe I}/N_{\rm Fe II}$
varies between $0.01$ and $3.0$ depending on optical depth, metallicity and
gravity. Here, photoionization can have an effect, although the radiative rates
are generally small due to the low $\Teff$ of the models in combination with the
extreme line blanketing (recall that the metallicities are close to solar).

The analysis of the \fei\ transition rates showed that for all models with
$\Teff = 3400$ to $3800$ K, irrespective of their $\log g$ and [Fe/H],
collisions fully dominate over radiatively-induced transitions out to the depths
$\opd \sim -3$ (Fig. \ref{dep1}, top panel). Only near the outer boundary, $\opd
\leq -3$, the departure coefficients of the upper level of the multiplet $296$,
\Fe{z}{5}{D}{\circ}{}, deviate from unity, whereas the lower level
\Fe{a}{5}{P}{}{} retains its LTE populations all over the optical depth scale.
As a result, deviations from LTE in the opacity of the \fei\ lines are
negligible, and the NLTE and LTE line formation depths nearly coincide (as shown
by the vertical marks in Fig. \ref{dep1}). Very small NLTE effects, which show
up in the line cores, are entirely due to the deviation of the line source
function $S_{\rm ij}$ from the Planck function $B_{\nu}(T_{\rm e})$. Since
$S_{\rm ij}/B_{\nu}(T_{\rm e}) \sim b_j/b_i < 1$\footnote{Hereafter, $i$ and $j$
subscripts stand for the lower and upper level, respectively, and b$_{\rm i}$,
b$_{\rm j}$ are the corresponding departure coefficients, i.e., the ratio of
NLTE to LTE occupation numbers.}, that is driven by the photon escape in the
line wings, the line cores are slightly darker under NLTE and the NLTE abundance
corrections $\Delta_{\rm Fe}$ are negative (Table 2 and 3). Metallicity and
surface gravity have very small influence on the magnitude of NLTE effects for
these cool models.

\begin{figure*}[ht!]
\begin{center}
\resizebox{0.8\textwidth}{!}{\includegraphics[scale=1]
{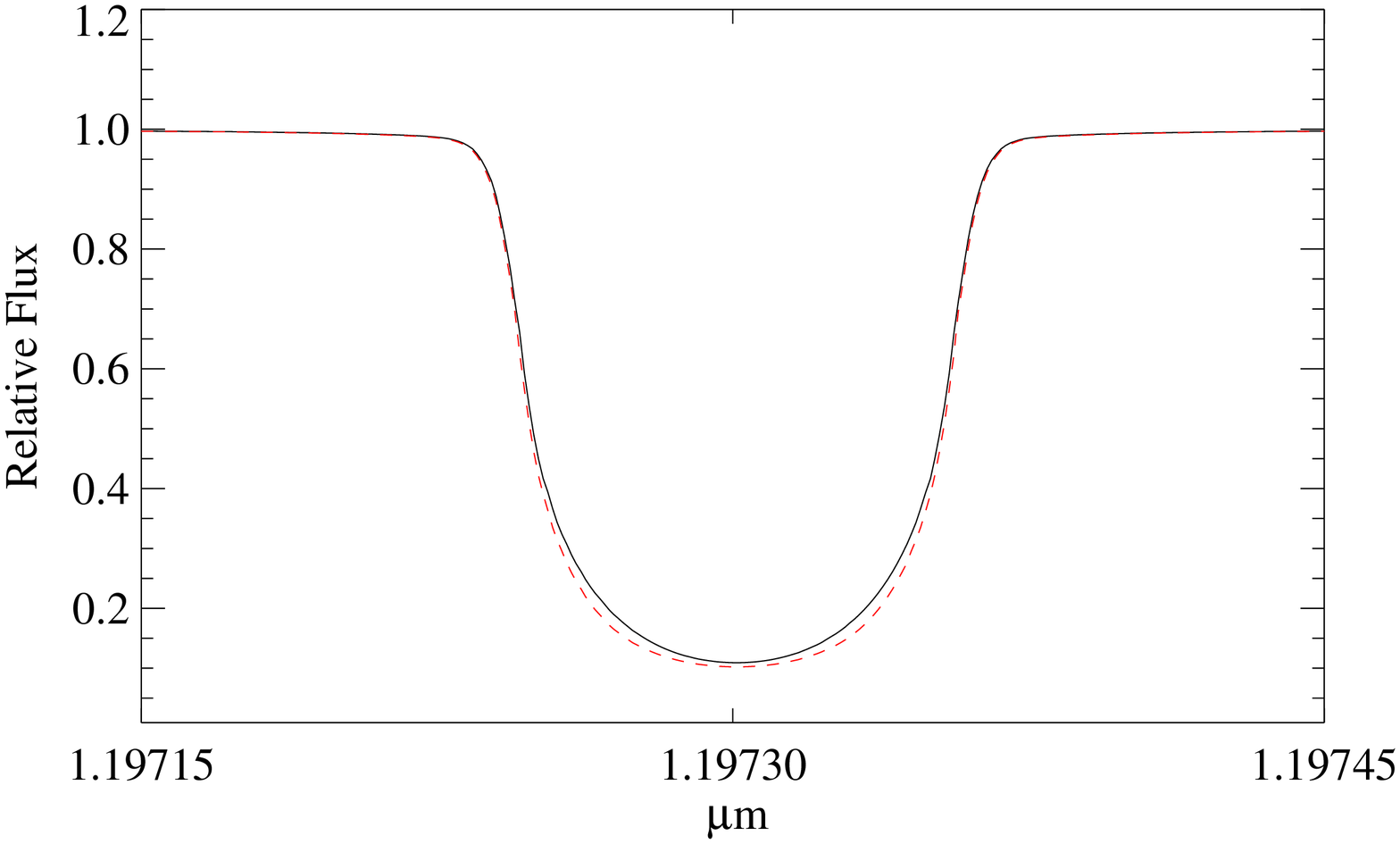}}
\caption{The NLTE (black, solid) and LTE (red, dashed) profiles of the Fe I line
at $11973$ \AA\ computed for T$_{\rm eff}$ = 4400K, log g = -0.5 and [Z] = -0.5
with microturbulence $\xi_{\rm t} = 5$ km/s.}
\label{prof_fe}
\end{center}
\end{figure*}

Larger departures from LTE are found for the hottest models with $\Teff$ =
4400K (Fig. \ref{dep1}, bottom panel). They are caused by radiative pumping from
the lower \Fe{a}{5}{P}{}{} term to higher levels and strong collisional
coupling. As a result \Fe{a}{5}{P}{}{} becomes progressively more underpopulated
in the inner atmospheric layers and the line absorption coefficient is reduced.
At the same time collisional coupling induces $b_i \sim b_j$ and the line source
function remains at the LTE value. As a consequence, the NLTE profiles become
brighter in the core than the LTE profiles and the NLTE equivalent widths are
smaller in NLTE. The magnitude of the NLTE effects decreases with increasing
gravity and metallicity. The most extreme effects are found at $\Teff= 4400$ K,
$\log g =-0.5$, and [Fe/H] $= -0.5$. NLTE and LTE line profiles for these
parameters and for the model atmosphere with $\Teff= 4400$ K, $\log g =-0.5$,
and [Fe/H] $= 0.5$, are shown in Fig. \ref{prof_fe}. The value of the
micro-turbulence has almost no influence on the size of the NLTE effects.
\subsection{Statistical equilibrium of Ti}

For \tii, our calculations show that non-LTE effects are more significant than
those for \fei\ discussed in the previous section. The atomic structure of \tii\
(ground state electronic configuration $1s^2 2s^2 2p^6 3s^2 3p^6 3d^2 4s^2$) is
simpler than that of \fei\ . In the NLTE model of \fei\ , a very large number of
levels allow for a more efficient collisional and radiative coupling, generally
leading to a stronger overall thermalization of the level populations. We also
note that the ionization potential of \tii\ is only $6.82$ eV ($1.1$ eV lower
than that of \fei), thus the ionization balance is such that \tiii\ is the
dominant ionization stage for all models in our grid. The ratio of $N_{\rm Ti
I}/N_{\rm Ti II}$ is 10$^{-2}$ to 10$^{-3}$ for the 4400K models and $0.1$ to 1
for the 3400K models.

\begin{figure*}[ht!]
\begin{center}
\resizebox{0.75\textwidth}{!}{\includegraphics[scale=0.8]
{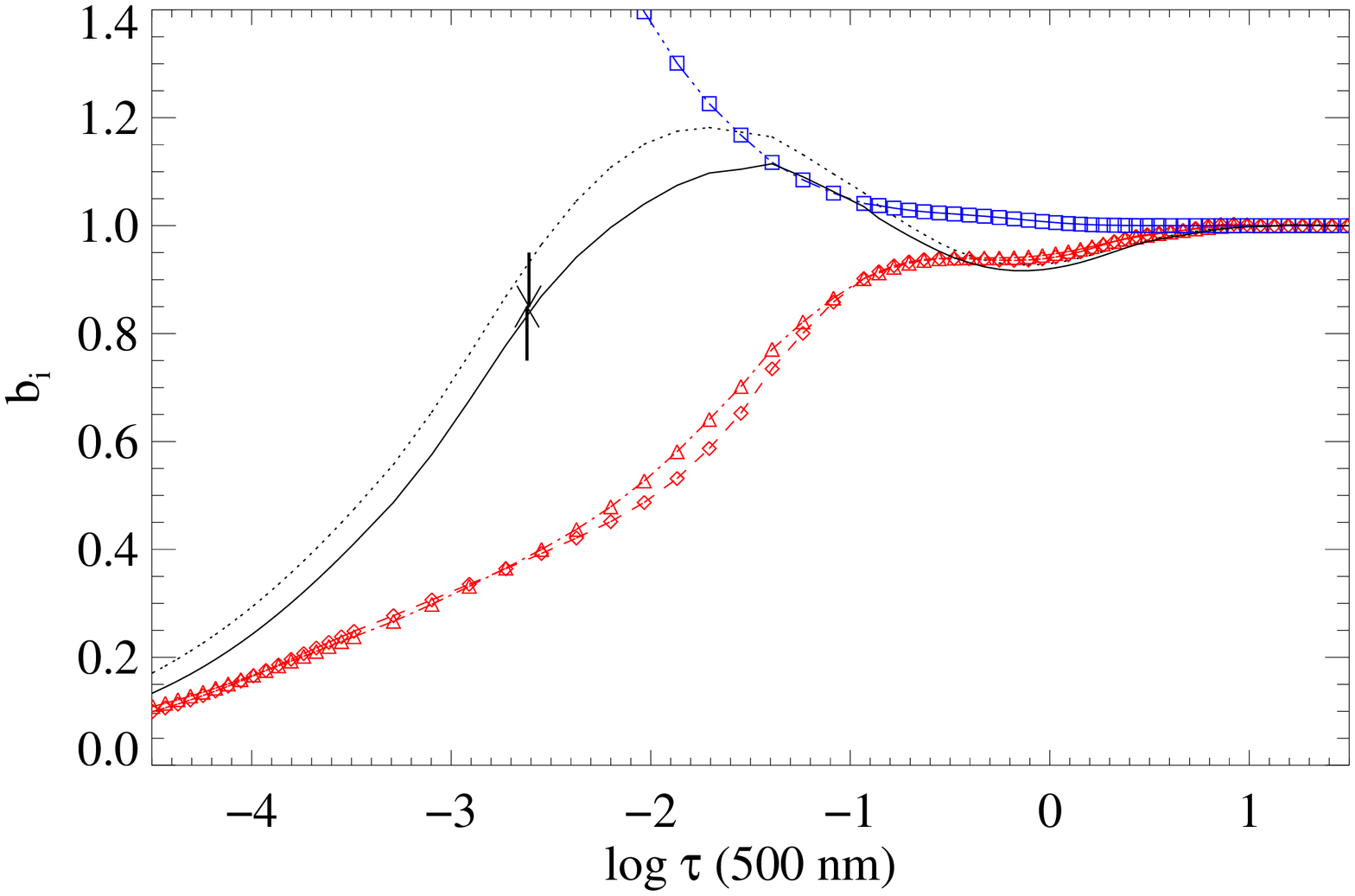}}
\resizebox{0.75\textwidth}{!}{\includegraphics[scale=0.8]
{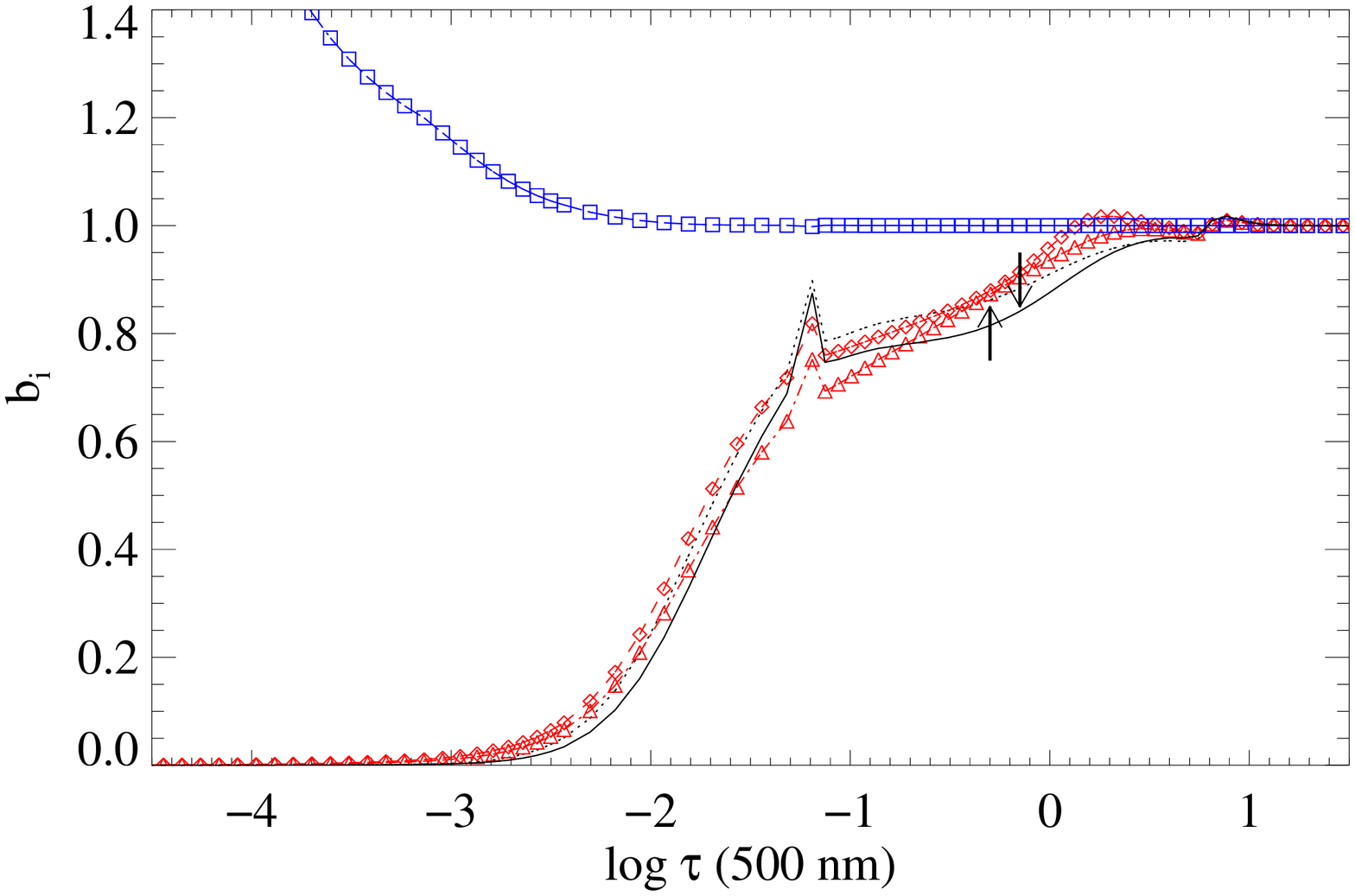}}
\caption{The NLTE departure coefficients of \tii\ for RSG models with log g =
-0.5, [Z] = -0.5 and T$_{\rm eff}$ = 3400K (top) and 4400K (bottom).
Blue: \tiii\ ground state \Ti{a}{4}F{}{}. Black: lower fine structure levels 
of J-band IR transitions, \Ti{b}{3}F{}{3} (dotted), \Ti{b}{3}F{}{2} (solid). 
Red: upper fine structure levels of IR-transitions, \Ti{z}{3}D{\circ}{2}
(triangles), \Ti{z}{3}D{\circ}{1} (diamonds). The LTE and NLTE line center
optical depths $\log \tau (11949.58$ \AA, \tii$) = 0$ are also indicated by the
upward and downward directed arrows, respectively.}
\label{dep2}
\end{center}
\end{figure*}

\begin{figure*}[ht!]
\begin{center}
\resizebox{0.8\textwidth}{!}{\includegraphics[scale=1]
{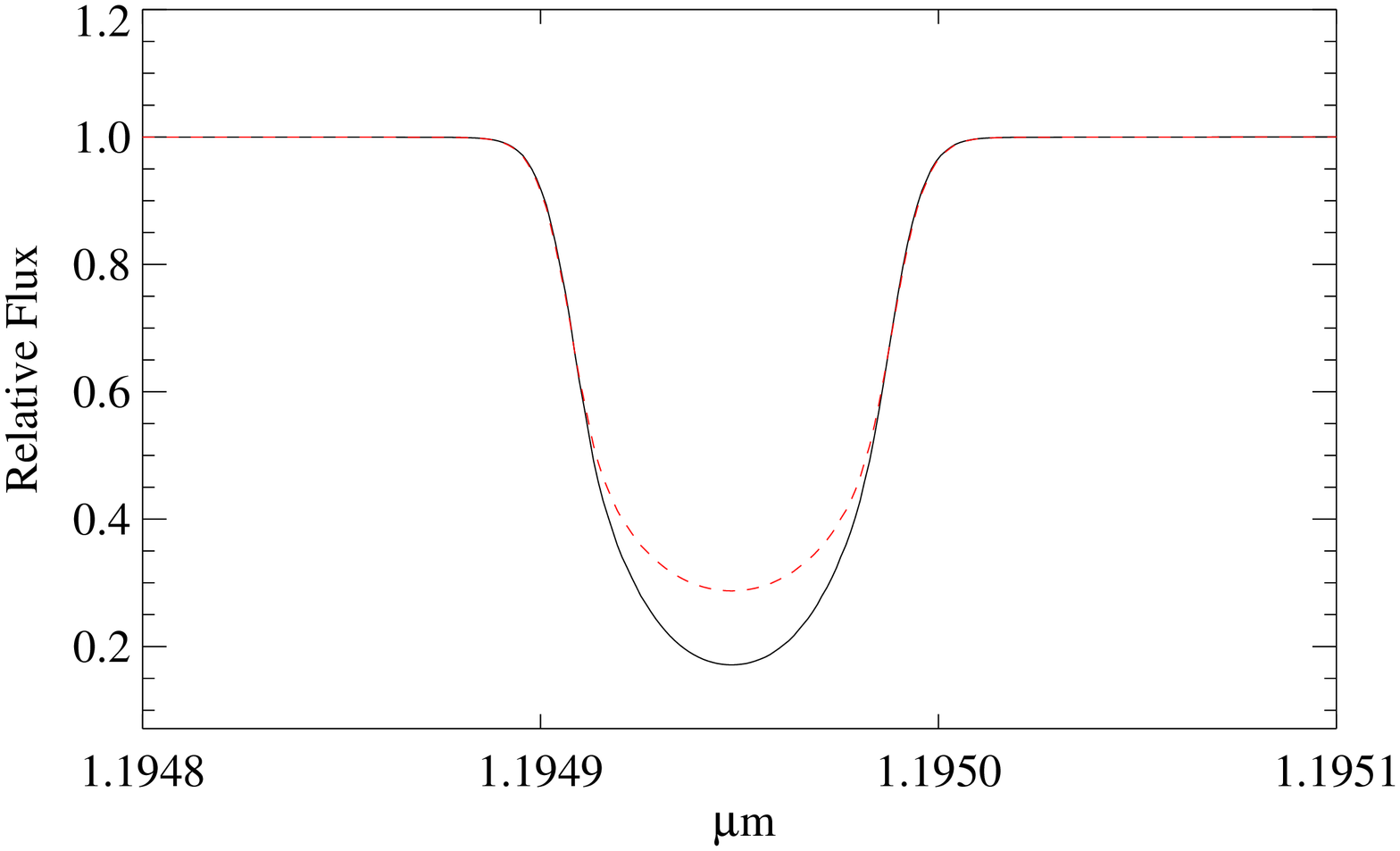}}
\resizebox{0.8\textwidth}{!}{\includegraphics[scale=1]
{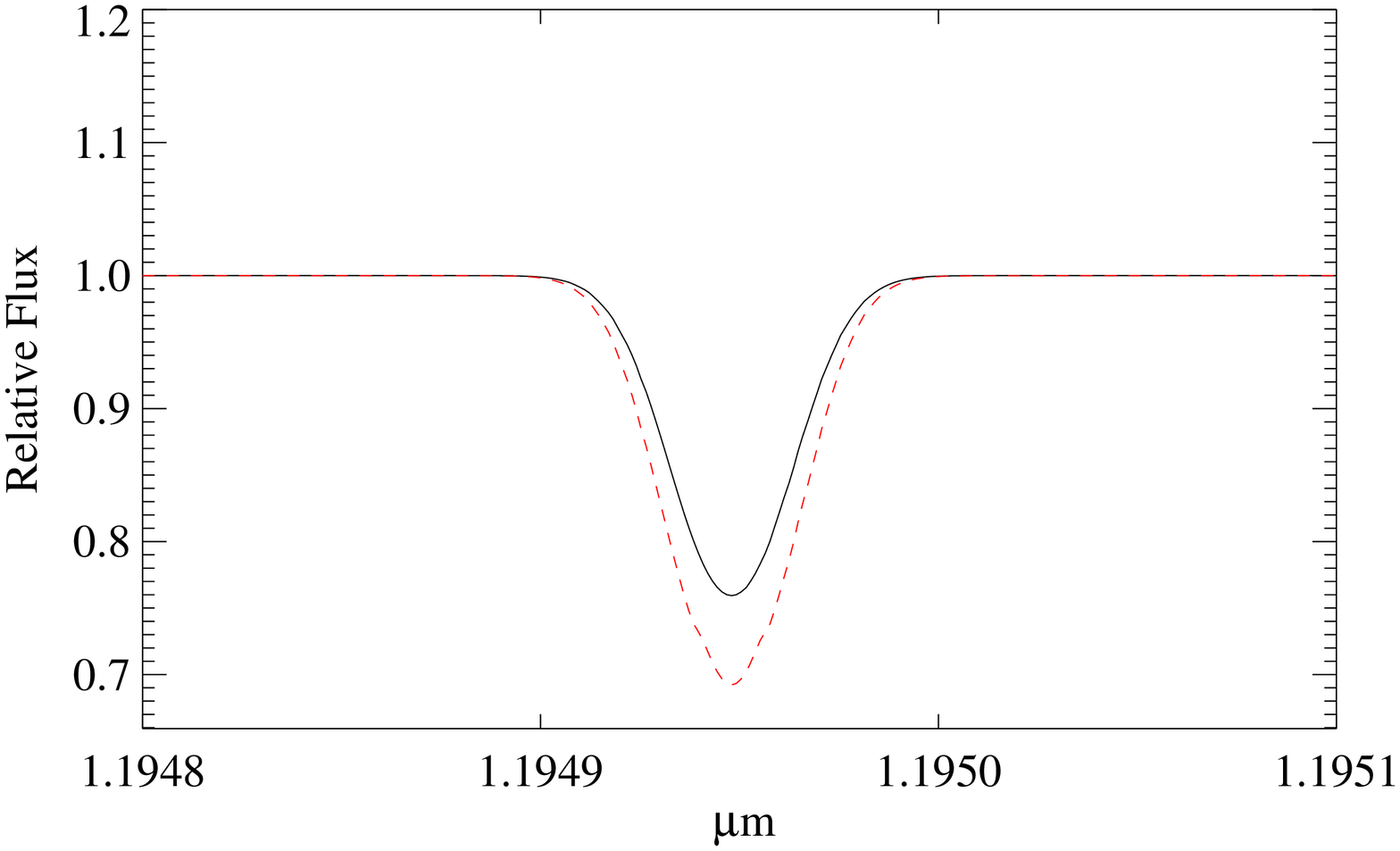}}
\caption{The NLTE (black, solid) and LTE (red, dashed) profiles of the \tii line
at $11949.58$ \AA\ computed for log g = -0.5 and [Z] = -0.5 with microturbulence
$\xi_{\rm t} = 5$ km/s and T$_{\rm eff}$ = 3400K (top) and 4400K (bottom).}
\label{prof_ti}
\end{center}
\end{figure*}

The departure coefficients for the \Ti{b}{3}{F}{}{} and \Ti{z}{3}{D}{\circ}{}
levels are shown in the Fig. \ref{dep2}. These levels with excitation energies
of $1.43$ respectively $2.47$ eV give rise to the diagnostic IR \tii\
transitions (Table 1). The \tii\ NLTE effects are a strong function of effective
temperature. At T$_{\rm eff}$ = 3400K the lower (metastable) \Ti{b}{3}{F}{}{}
states are overpopulated in the region of line formation through radiative
transitions from higher levels and the upper levels \Ti{z}{3}{D}{\circ}{} are
depleted by radiative transitions to lower levels, in particular to the ground
state. In consequence, the  \tii\ J-band absorption lines become stronger in
NLTE compared to LTE because of the enhanced line absorption coefficient and the
sub-thermal source function (see Fig. \ref{prof_ti}).

The situation changes towards higher effective temperatures. At T$_{\rm eff}$ =
4400K the radiation field has become powerful enough to deplete the lower
\Ti{b}{3}{F}{}{} states through radiative pumping into higher levels. While the
upper \Ti{z}{3}{D}{\circ}{} are populated from below in this way, they are also
de-populated by radiative pumping to even higher levels, which leads to a net
underpopulation. The strong de-population of the lower levels weakens the line
absorption coefficient, while the source function remains close to the LTE
value. As a result, the \tii\ absorption lines become weaker in NLTE at higher
temperatures (see Fig. \ref{prof_ti}).

\subsection{Equivalent widths and non-LTE abundance corrections} 

\begin{figure*}[ht!]
\begin{center}
\includegraphics[scale=0.60]{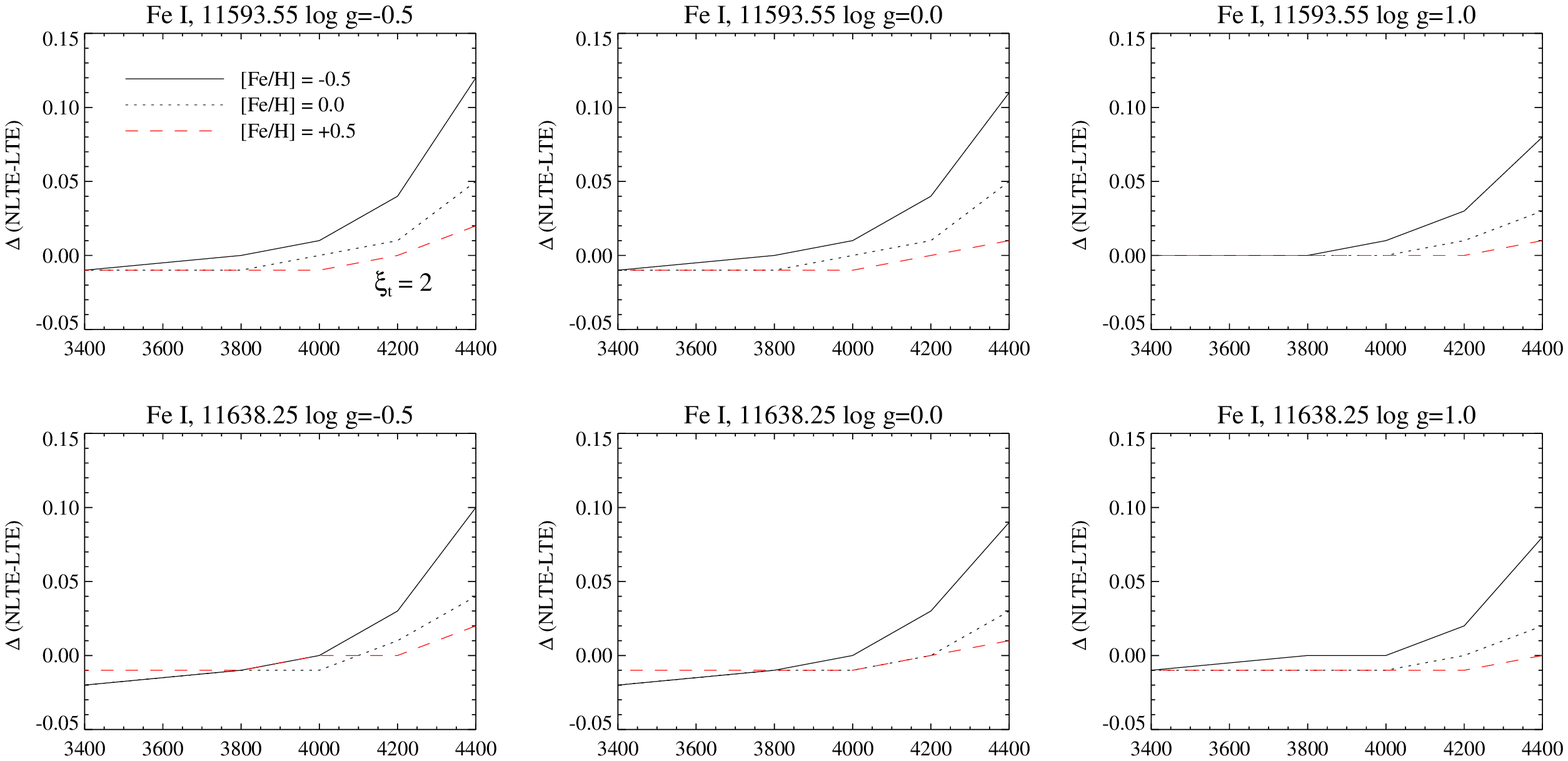}
\includegraphics[scale=0.60]{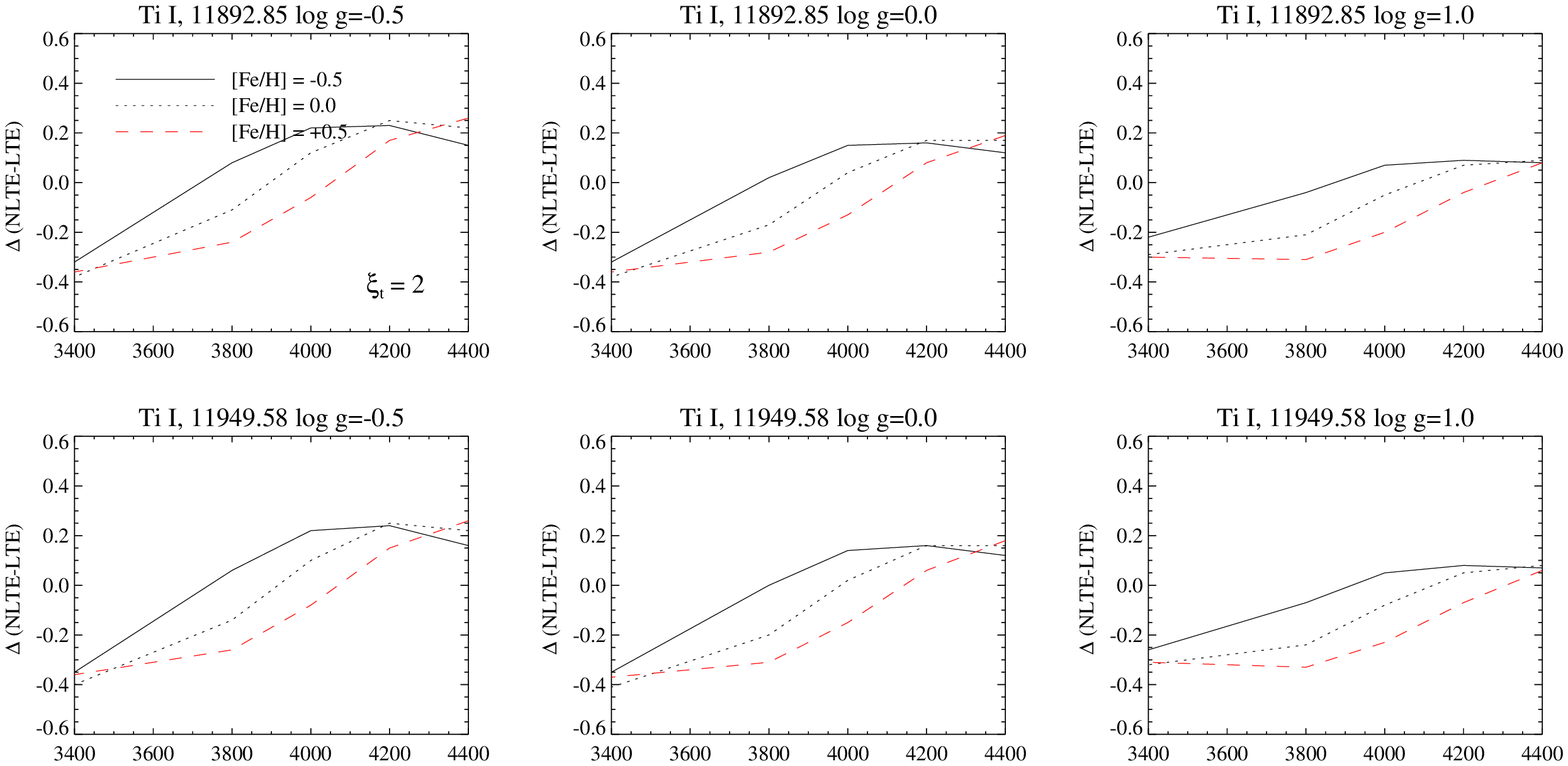}
\caption{NLTE abundance corrections as a function of effective temperature for 
microturbulence $\xi$ = 2 km/s for \fei 11593 \AA (top),\fei 11638 \AA (2nd
row), \tii 11893 \AA (3rd row) and \tii 11949 (bottom). Left column: log g =
-0.5, middle column: log g = 0.0, right column: log g = 1.0. Black solid: [Z] =
-0.5, blue dotted: [Z] = 0.0, red dashed: [Z] = +0.5.}
\label{nltecor2}
\end{center}
\end{figure*}

The NLTE and LTE abundance corrections and equivalent widths for the 
individual \tii\ and \fei\ J-band lines studied are given in Tables 2 to 5.
Fig. \ref{nltecor2} shows the NLTE abundance corrections for some of
these lines computed using $\xi = 2$ km s$^{-1}$ as a function of effective
temperature, gravity and metallicity. The results for microturbulence 5 km
s$^{-1}$ are not shown because they are nearly identical. As a consequence of
the small \fei\ NLTE effects discussed in section 4.1 the NLTE abundance
corrections are very small and reach a maximum value $\sim$ 0.1 dex only at the
highest effective temperature. The medium resolution J-band $\chi^{2}$ fitting
method of RSG spectra developed by DKF10 achieves an accuracy of $0.15$ dex on
average for the metallicity of an individual RSG. Thus, we do not expect NLTE
effects in \fei\ to heavily affect the results.

On the other hand, the NLTE effects for \tii\ are more pronounced as discussed
in section 3.2. This is clearly reflected in the NLTE abundance corrections,
which become $\sim -0.4$ dex at low effective temperatures and change to $\sim
0.2$ dex at high effective temperature. Thus, if the \tii\ lines were the only
features to derive stellar metallicities from J-band spectra, NLTE effects would
imply large corrections. However, the technique introduced by DKF10 uses the
full information of many J-band lines from different atomic species (7 \fei, 2
\mgi, 2 \tii, 3 \sii\ lines) simultaneously to determine metallicity together
with stellar parameters. Since the \fei\ lines are only weakly affected by NLTE,
we expect \tii\ NLTE effects to have a smaller influence on the determination of
the overall metallicity than the NLTE abundance corrections for the \tii\
near-IR lines only.

\subsection{J-band medium resolution spectral analysis}

In order to assess the influence of the NLTE effects on the J-band medium 
resolution metallicity studies, we have carried out the following experiment.
We calculated complete synthetic J-band spectra with MARCS model atmospheres and
LTE opacities for all spectral lines except the \fei\ and \tii\, for which we
used our NLTE calculations. We then used these synthetic spectra calculated for
T$_{\rm eff}$ from 3400K to 4400K with different log g and different metallicity
(and with added Gaussian noise corresponding to S/N of $200$) as input for the
DKF10 $\chi^{2}$ analysis using MARCS model spectra calculated completely in
LTE. The metallicity grid for the MARCS model spectra had a resolution of 0.25
dex between [Z] = -1 and +0.5 and a cubic spline interpolation was used between
the grid points (see \citealt{evans11}). From the metallicities recovered in
this way we can estimate the possible systematic errors when relying on a
complete LTE fit of RSG J-band spectra.

We found small average metallicity corrections of $-0.15$ dex at low
temperatures and $+0.1$ dex at high temperatures with only a weak dependence on
input metallicity (Fig. 11). This qualitative behaviour is as expected given the
results shown in Fig. 9 and 10. Quantitatively, the effects are small, since the
results of these tests are dominated by the \fei\ lines, which are more numerous
than the \tii\ lines and for which the behavior NLTE corrections are minor (note
that in this paper we have only discussed the non-LTE effects of the four
strongest \fei\ lines). The [Z] resolution of the model grid
for this experiment is 0.25 dex and we know that the systematics uncertainties
of the analysis technique are of the order of 0.15 dex (DKF10), thus, the NLTE
corrections are marginally within the accuracy of the fitting procedure.
However, very obviously the inclusion of NLTE effects in the calculation of
J-band \fei\ and, in particular, \tii\ will improve the accuracy of future work.

\begin{figure*}[ht!]
\begin{center}
\resizebox{0.95\textwidth}{!}{\includegraphics[scale=0.5]
{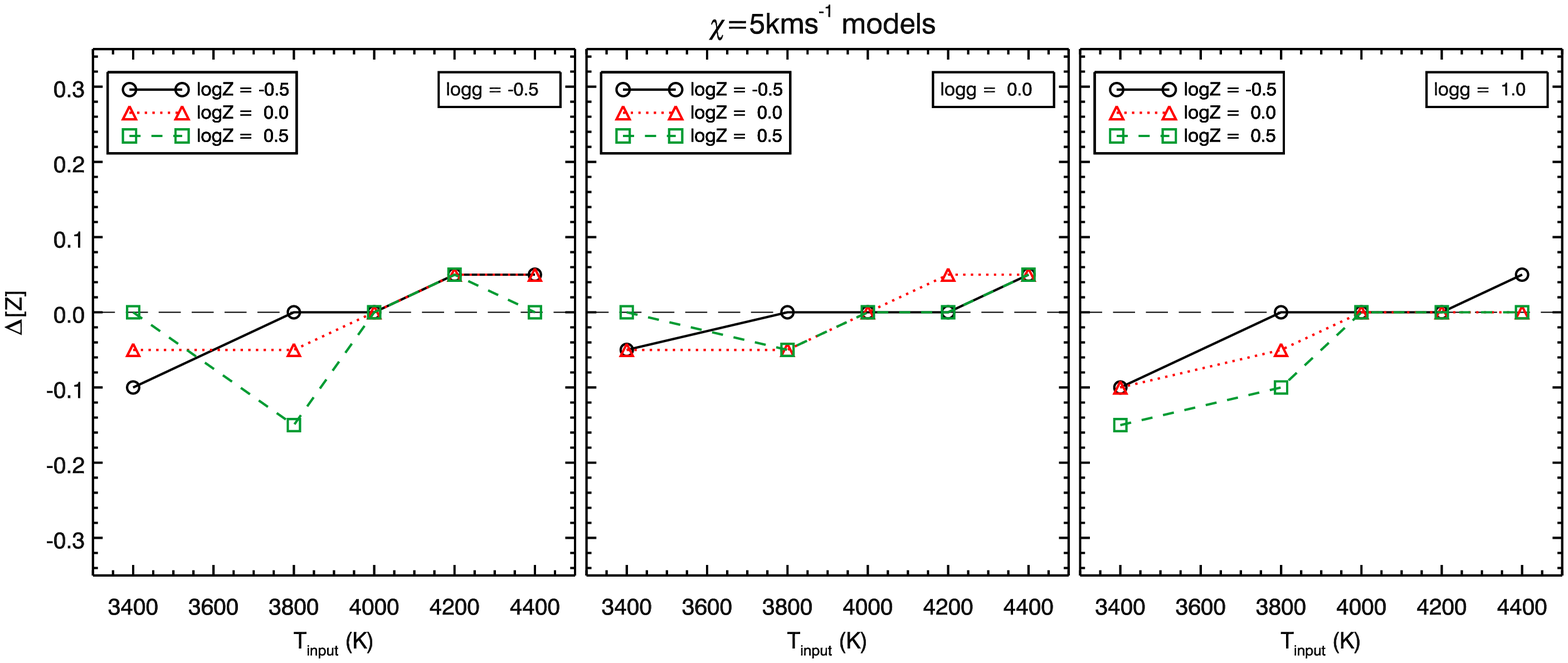}}
\caption{Influence of the TiI and FeI non-LTE effects on the DKF10 J-band
$\chi^{2}$ 
metallicity determination as a function of effective temperature. The numerical 
experiment is described in the text. Left: log g $= -0.5$, middle: log g $=
0.0$, right: log g $= 1.0$. Circles: [Z] $= -0.5$, triangles: [Z] $= 0.0$,
squares: [Z] $= 0.5$
}
\label{nlteparam}
\end{center}
\end{figure*}

\subsection{Modeling uncertainties}

Spectroscopic analysis of red supergiants is generally a highly
non-trivial matter \citep{1992iesh.conf...86G}, as other complexities arise.
Among the most severe is sub-photospheric convection. At present, it is not
possible to systematically evaluate the effect of 1D hydrostatic equilibrium
approximation on the determination of stellar parameters from RSG spectra. The
only realistic non-gray 3D radiative hydrodynamics simulation of convection for
a single RSG model was performed by \citet{2011A&A...535A..22C}. They
demonstrated that there is no unique 1D static model, which would provide
similar observable characteristics, i.e. SED shape and colors, as the analogous
3D RHD model. To match the optical spectrum dominated by TiO opacity, a $\sim
100$ K cooler 1D model is needed, whereas the IR spectral window can be best
matched using a $\sim 200$ K hotter 1D model. The NLTE effects, however, are
sensitive to the radiation field across the whole spectrum, from UV to the IR.
Thus, it is not possible to predict the changes in the NLTE abundance
corrections simply by using a cooler or hotter 1D static model. One could, in
principle, expect that, since the mean T($\tau)$ relation of a typical 3D RHD
model \citep[][their Fig. 5]{2011A&A...535A..22C} is shallower than that of a
corresponding 1D static model, the NLTE effects would be smaller. Our recent
work on low-mass solar-like stars \citep{bergemann12} showed that the largest
differences in terms of NLTE effects occur at low metallicities, where 3D
models are up to $1000$ K cooler than 1D hydrostatic models. The RSG models
investigated in this work have metallicities close to solar. In any case, full
NLTE calculations adopting, at least, a mean structure of the 3D RHD model, are
necessary. Given our recent experience with calculations of such type for the
solar-like stars (Bergemann et al. 2012), we are now in position to perform
similar analysis for RSG's.

\section{Conclusions and Future Work}

We have presented detailed calculations of the non-LTE effects in
\fei\ and \tii\ lines in the J-band, relevant to the analysis technique of
DFK10.  The non-LTE corrections to abundances measured from individual \fei\ 
lines are small -- less than $0.15$ dex across the parameter range studied. 
On the other hand, for \tii\ these corrections are larger, $0.4$ dex
at the extreme. However, as there are only two \tii\ lines in the J-band
spectral window compared to many \fei\ lines and other lines of other 
elements such as \mgi\ and \sii\, the impact on the derived metallicity of a
given spectrum is small. Differential analysis of non-LTE corrected and
uncorrected spectra shows that the average difference in measured metallicity
$[Z]$ is $\la$ 0.15 dex, i.e. still within the accuracy range of the medium
resolution analysis procedure but large enough and showing systematic trends
with $T_{\rm eff}$ and $[Z]$ that it seems worthwhile to include the non-LTE
effects in the calculation of the J-band synthetic spectra. This will ultimately
improve the accuracy of the method, in particular, when it will be applied not
only to measure overall metallicity but also individual element abundances and
the diagnostically important ratio of $\alpha$ to iron elements.

In terms of future work, we will next turn our attention to the other
prominent species in the $J$-band spectral window, namely those of \sii\
and \mgi. These lines also contain important metallicity information,
since they help constrain the measurement of a star's $\alpha$-element
abundance. In order to determine the non-LTE corrections to the
$J$-band transitions of these species we have already begun to construct model
atoms, and the computation of the NLTE line strengths will begin shortly.


\acknowledgments
This work was supported by the National Science Foundation under grant
AST-1108906 to RPK. Moreover, RPK acknowledges support by the
Alexander-von-Humboldt Foundation and the hospitality  of the
Max-Planck-Institute for Astrophysics in Garching and the University
Observatory Munich, where part of this work was carried out.

\clearpage

\clearpage


\input{tab1}
\clearpage

\input{stab2}
\clearpage

\input{stab3}
\clearpage

\input{stab4}
\clearpage

\input{stab5}
\clearpage



\end{document}

%% file: tab1.tex

\begin{deluxetable}{ccccccc}
\tabletypesize{\scriptsize}
\tablecolumns{7}
\tablewidth{0pt}
\tablecaption{J-band \fei\ and \tii\ lines}

\tablehead{
\colhead{Elem.}     &
\colhead{$\lambda$}            &
\colhead{$\Elow$}           &
\colhead{lower}     &
\colhead{$\Eup$}     &
\colhead{upper}             &
\colhead{$log~gf$}\\
\colhead{}     &
\colhead{\AA}            &
\colhead{[eV]}           &
\colhead{level}     &
\colhead{[eV]}     &
\colhead{level}             &
\colhead{}\\[1mm]
\colhead{(1)}	&
\colhead{(2)}	&
\colhead{(3)}	&
\colhead{(4)}	&
\colhead{(5)}	&
\colhead{(6)}	&
\colhead{(7)}}	
\startdata
\\[-1mm]
\tii & 11892.85 & 1.43 & \Ti{b}{3}{F}{}{2} & 2.47 & \Ti{z}{3}{D}{\circ}{1} & -1.908\\ 
            & 11949.58 & 1.44 & \Ti{b}{3}{F}{}{3} & 2.48 & \Ti{z}{3}{D}{\circ}{2} & -1.760\\  
\fei & 11638.25 & 2.18 & \Fe{a}{5}{P}{}{3} & 3.25 & \Fe{z}{5}{D}{\circ}{3} & -2.214\\
            & 11973.01 & 2.18 & \Fe{a}{5}{P}{}{3} & 3.22 & \Fe{z}{5}{D}{\circ}{4} & -1.483\\ 
            & 11882.80 & 2.20 & \Fe{a}{5}{P}{}{2} & 3.24 & \Fe{z}{5}{D}{\circ}{3} & -1.668\\ 
            & 11593.55 & 2.22 & \Fe{a}{5}{P}{}{1} & 3.29 & \Fe{z}{5}{D}{\circ}{0} & -2.448\\
\enddata
\end{deluxetable}

%% file: stab2.tex
%
%

\begin{deluxetable}{ccccccccc}
\tabletypesize{\scriptsize}
\tablecolumns{9}
\tablewidth{0pt}
\tablecaption{Non-LTE abundance corrections for the \tii\ and \fei\ lines ($\xi$ = 2 kms$^{-1}$)}

\tablehead{
\colhead{T$_{\rm eff}$}     &
\colhead{$log~g$}            &
\colhead{[Z]}           &
\colhead{$\Delta_{\rm Ti I}$}     &
\colhead{$\Delta_{\rm Ti I}$}             &
\colhead{$\Delta_{\rm Fe I}$}     &
\colhead{$\Delta_{\rm Fe I}$}             &
\colhead{$\Delta_{\rm Fe I}$}     &
\colhead{$\Delta_{\rm Fe I}$}\\
\colhead{}     &
\colhead{}            &
\colhead{}           &
\colhead{$11892.85$}     &
\colhead{$11949.58$}     &
\colhead{$11593.55$}             &
\colhead{$11638.25$}     &
\colhead{$11882.80$}             &
\colhead{$11973.01$}\\[1mm]
\colhead{(1)}	&
\colhead{(2)}	&
\colhead{(3)}	&
\colhead{(4)}	&
\colhead{(5)}	&
\colhead{(6)}	&
\colhead{(7)}   &
\colhead{(8)}	&
\colhead{(9)}}	
\startdata
\\[-1mm]
 4400. & -0.50 &  0.50 &  0.26  &  0.26 &  0.02 &  0.02 &  0.02 &  0.02 \\
 4400. & -0.50 &  0.00 &  0.22  &  0.22 &  0.05 &  0.04 &  0.03 &  0.03 \\
 4400. & -0.50 & -0.50 &  0.15  &  0.16 &  0.12 &  0.10 &  0.08 &  0.07 \\
 4400. &  0.00 &  0.50 &  0.19  &  0.18 &  0.01 &  0.01 &  0.01 &  0.01 \\
 4400. &  0.00 &  0.00 &  0.17  &  0.16 &  0.05 &  0.03 &  0.02 &  0.02 \\
 4400. &  0.00 & -0.50 &  0.12  &  0.12 &  0.11 &  0.09 &  0.07 &  0.06 \\
 4400. &  1.00 &  0.50 &  0.08  &  0.06 &  0.01 &  0.00 & -0.01 & -0.01 \\
 4400. &  1.00 &  0.00 &  0.09  &  0.08 &  0.03 &  0.02 &  0.01 &  0.01 \\
 4400. &  1.00 & -0.50 &  0.08  &  0.07 &  0.08 &  0.08 &  0.05 &  0.05 \\
 & & & & & & & & \\ 
\enddata
\tablecomments{This table is published in its entirety in the electronic edition
of ApJ. A portion is shown here for guidance regarding its form and content.}
\end{deluxetable}

%% file: stab3.tex
%
%

\begin{deluxetable}{ccccccccc}
\tabletypesize{\scriptsize}
\tablecolumns{9}
\tablewidth{0pt}
\tablecaption{Non-LTE abundance corrections for the \tii\ and \fei\ lines ($\xi$ = 5 kms$^{-1}$)}

\tablehead{
\colhead{T$_{\rm eff}$}     &
\colhead{$log~g$}            &
\colhead{[Z]}           &
\colhead{$\Delta_{\rm Ti I}$}     &
\colhead{$\Delta_{\rm Ti I}$}             &
\colhead{$\Delta_{\rm Fe I}$}     &
\colhead{$\Delta_{\rm Fe I}$}             &
\colhead{$\Delta_{\rm Fe I}$}     &
\colhead{$\Delta_{\rm Fe I}$}\\
\colhead{}     &
\colhead{}            &
\colhead{}           &
\colhead{$11892.85$}     &
\colhead{$11949.58$}     &
\colhead{$11593.55$}             &
\colhead{$11638.25$}     &
\colhead{$11882.80$}             &
\colhead{$11973.01$}\\[1mm]
\colhead{(1)}	&
\colhead{(2)}	&
\colhead{(3)}	&
\colhead{(4)}	&
\colhead{(5)}	&
\colhead{(6)}	&
\colhead{(7)}   &
\colhead{(8)}	&
\colhead{(9)}}	
\startdata
\\[-1mm]
 4400. & -0.50 &  0.00 &  0.25 &  0.25 &  0.03 &  0.03 &  0.03 &  0.03 \\
 4400. & -0.50 &  0.00 &  0.22 &  0.22 &  0.06 &  0.06 &  0.04 &  0.04 \\
 4400. & -0.50 & -0.50 &  0.17 &  0.16 &  0.12 &  0.12 &  0.11 &  0.11 \\
 4400. &  0.00 &  0.50 &  0.20 &  0.20 &  0.02 &  0.01 &  0.00 &  0.01 \\
 4400. &  0.00 &  0.00 &  0.17 &  0.17 &  0.06 &  0.05 &  0.03 &  0.03 \\
 4400. &  0.00 & -0.50 &  0.15 &  0.14 &  0.11 &  0.11 &  0.10 &  0.10 \\
 4400. &  1.00 &  0.50 &  0.11 &  0.10 &  0.01 &  0.00 & -0.02 & -0.02 \\
 4400. &  1.00 &  0.00 &  0.11 &  0.11 &  0.04 &  0.03 &  0.02 &  0.01 \\
 4400. &  1.00 & -0.50 &  0.12 &  0.10 &  0.08 &  0.09 &  0.08 &  0.08 \\
 & & & & & & & & \\
\enddata
\tablecomments{This table is published in its entirety in the electronic edition
of ApJ. A portion is shown here for guidance regarding its form and content.}
\end{deluxetable}

%% file: stab4.tex
%
%

\begin{deluxetable}{ccccccccccccccc}
\tabletypesize{\scriptsize}
\tablecolumns{15}
\tablewidth{0pt}
\tablecaption{Equivalent widths \tablenotemark{a}  of the \tii\ and \fei\ lines ($\xi$ = 2 kms$^{-1}$)}

\tablehead{
\colhead{T$_{\rm eff}$}     &
\colhead{$log~g$}            &
\colhead{[Z]}           &
\colhead{$W_{\lambda,\rm Ti I}$}     &
\colhead{$W_{\lambda,\rm Ti I}$}             &
\colhead{$W_{\lambda,\rm Ti I}$}     &
\colhead{$W_{\lambda,\rm Ti I}$}             &
\colhead{$W_{\lambda,\rm Fe I}$}     &
\colhead{$W_{\lambda,\rm Fe I}$}             &
\colhead{$W_{\lambda,\rm Fe I}$}     &
\colhead{$W_{\lambda,\rm Fe I}$}             &
\colhead{$W_{\lambda,\rm Fe I}$}     &
\colhead{$W_{\lambda,\rm Fe I}$}             &
\colhead{$W_{\lambda,\rm Fe I}$}     &
\colhead{$W_{\lambda,\rm Fe I}$}\\
\colhead{}     &
\colhead{}            &
\colhead{}           &
\colhead{$11892$}     &
\colhead{$11892$}     &
\colhead{$11949$}     &
\colhead{$11949$}     &
\colhead{$11593$}             &
\colhead{$11593$}             &
\colhead{$11638$}     &
\colhead{$11638$}     &
\colhead{$11882$}             &
\colhead{$11882$}             &
\colhead{$11882$}             &
\colhead{$11973$}\\[1mm]
\colhead{}     &
\colhead{}            &
\colhead{}           &
\colhead{$LTE$}     &
\colhead{$NLTE$}     &
\colhead{$LTE$}     &
\colhead{$NLTE$}     &
\colhead{$LTE$}             &
\colhead{$NLTE$}             &
\colhead{$LTE$}     &
\colhead{$NLTE$}     &
\colhead{$LTE$}             &
\colhead{$NLTE$}             &
\colhead{$LTE$}             &
\colhead{$NLTE$}\\[1mm]
\colhead{(1)}	&
\colhead{(2)}	&
\colhead{(3)}	&
\colhead{(4)}	&
\colhead{(5)}	&
\colhead{(6)}	&
\colhead{(7)}   &
\colhead{(8)}	&
\colhead{(9)}	&
\colhead{(10)}   &
\colhead{(11)}	&
\colhead{(12)}	&
\colhead{(13)}   &
\colhead{(14)}	&
\colhead{(15)}}	
\startdata
\\[-1mm]
4400. & -0.50 &  0.50 & 221.6 & 186.3 & 239.7 & 205.6 & 438.5 & 436.0 & 492.1 & 488.2 & 658.1 & 649.8 &  773.3 &  759.5 \\    
4400. & -0.50 &  0.00 & 155.8 & 123.2 & 176.2 & 143.3 & 386.5 & 380.7 & 425.9 & 420.4 & 530.0 & 522.5 &  597.4 &  586.5 \\
4400. & -0.50 & -0.50 &  80.1 &  61.4 &  98.6 &  77.8 & 336.4 & 324.9 & 370.2 & 358.9 & 443.0 & 431.8 &  483.4 &  470.6 \\
4400. &  0.00 &  0.50 & 218.6 & 194.5 & 236.1 & 213.6 & 428.4 & 426.6 & 482.0 & 480.2 & 647.6 & 643.7 &  762.2 &  754.5 \\
4400. &  0.00 &  0.00 & 149.4 & 126.2 & 168.7 & 146.1 & 373.7 & 368.4 & 413.3 & 408.8 & 518.2 & 512.6 &  585.8 &  577.6 \\
& & & & & & & & & & & & & & \\
\enddata
\tablenotetext{a}{equivalent widths $\EW$ are given in \mA}
\tablecomments{This table is published in its entirety in the electronic edition
of ApJ. A portion is shown here for guidance regarding its form and content.}
\end{deluxetable}

%% file: stab5.tex
%
%

\begin{deluxetable}{ccccccccccccccc}
\tabletypesize{\scriptsize}
\tablecolumns{15}
\tablewidth{0pt}
\tablecaption{Equivalent widths \tablenotemark{a}  of the \tii\ and \fei\ lines ($\xi$ = 5 kms$^{-1}$)}

\tablehead{
\colhead{T$_{\rm eff}$}     &
\colhead{$log~g$}            &
\colhead{[Z]}           &
\colhead{$W_{\lambda,\rm Ti I}$}     &
\colhead{$W_{\lambda,\rm Ti I}$}             &
\colhead{$W_{\lambda,\rm Ti I}$}     &
\colhead{$W_{\lambda,\rm Ti I}$}             &
\colhead{$W_{\lambda,\rm Fe I}$}     &
\colhead{$W_{\lambda,\rm Fe I}$}             &
\colhead{$W_{\lambda,\rm Fe I}$}     &
\colhead{$W_{\lambda,\rm Fe I}$}             &
\colhead{$W_{\lambda,\rm Fe I}$}     &
\colhead{$W_{\lambda,\rm Fe I}$}             &
\colhead{$W_{\lambda,\rm Fe I}$}     &
\colhead{$W_{\lambda,\rm Fe I}$}\\
\colhead{}     &
\colhead{}            &
\colhead{}           &
\colhead{$11892$}     &
\colhead{$11892$}     &
\colhead{$11949$}     &
\colhead{$11949$}     &
\colhead{$11593$}             &
\colhead{$11593$}             &
\colhead{$11638$}     &
\colhead{$11638$}     &
\colhead{$11882$}             &
\colhead{$11882$}             &
\colhead{$11882$}             &
\colhead{$11973$}\\[1mm]
\colhead{}     &
\colhead{}            &
\colhead{}           &
\colhead{$LTE$}     &
\colhead{$NLTE$}     &
\colhead{$LTE$}     &
\colhead{$NLTE$}     &
\colhead{$LTE$}             &
\colhead{$NLTE$}             &
\colhead{$LTE$}     &
\colhead{$NLTE$}     &
\colhead{$LTE$}             &
\colhead{$NLTE$}             &
\colhead{$LTE$}             &
\colhead{$NLTE$}\\[1mm]
\colhead{(1)}	&
\colhead{(2)}	&
\colhead{(3)}	&
\colhead{(4)}	&
\colhead{(5)}	&
\colhead{(6)}	&
\colhead{(7)}   &
\colhead{(8)}	&
\colhead{(9)}	&
\colhead{(10)}   &
\colhead{(11)}	&
\colhead{(12)}	&
\colhead{(13)}   &
\colhead{(14)}	&
\colhead{(15)}}	
\startdata
\\[-1mm]

4400. & -0.50 &  0.50 & 387.6 & 304.7 & 432.6 & 349.8 & 871.8 & 866.0 &  944.3 &  938.2 & 1111.8 & 1106.5 & 1208.6 & 1198.9 \\
4400. & -0.50 &  0.00 & 234.0 & 170.0 & 278.6 & 209.9 & 781.9 & 768.3 &  850.8 &  838.4 &  990.4 &  980.0 & 1060.6 & 1047.5 \\
4400. & -0.50 & -0.50 &  96.0 &  68.2 & 123.1 &  91.3 & 678.3 & 651.4 &  748.6 &  722.2 &  877.9 &  853.4 &  936.3 &  910.1 \\
4400. &  0.00 &  0.50 & 385.6 & 324.5 & 428.4 & 369.3 & 847.7 & 843.4 &  920.2 &  917.0 & 1088.4 & 1087.5 & 1186.2 & 1181.5 \\
4400. &  0.00 &  0.00 & 225.4 & 176.5 & 268.0 & 216.7 & 751.8 & 739.6 &  820.4 &  809.8 &  960.5 &  952.4 & 1031.2 & 1020.8 \\
& & & & & & & & & & & & & & \\

\enddata
\tablenotetext{a}{equivalent widths $\EW$ are given in \mA}
\tablecomments{This table is published in its entirety in the electronic edition
of ApJ. A portion is shown here for guidance regarding its form and content.}
\end{deluxetable}

%% file: ms.bbl
\begin{thebibliography}{}
\bibitem[Allen(1973)]{1973asqu.book.....A} Allen, C.~W.\ 1973, London: 
University of London, Athlone Press, |c1973, 3rd ed.,  
\bibitem[Allende Prieto et al.(2001)]{allende01} Allende Prieto, C., Lambert,
D.~L., \& Asplund, M.\ 2001, \apjl, 556, L63 
\bibitem[Barklem et al.(2011)]{barklem11} Barklem, P.~S., Belyaev, A.~K.,
Guitou, M., et al.\ 2011, \aap, 530, A94
\bibitem[Barklem et al.(2012)]{barklem12} Barklem, P.~S., Belyaev, A.~K.,
Spielfiedel, A., Guitou, M., \& Feautrier, N.\ 2012, arXiv:1203.4877 
\bibitem[Bautista(1997)]{1997A&AS..122..167B} Bautista, M.~A.\ 1997, \aaps, 122,
167
\bibitem[Bergemann(2011)]{bergemann11} Bergemann, M.\ 2011, \mnras, 413, 2184
\bibitem[Bergemann et al.(2012)]{bergemann12} Bergemann, M., Lind, K., Collet,
R., Magic, Z., Asplund, M.\ 2012, MNRAS, submitted
\bibitem[Bresolin et al.(2009)]{bresolin09} Bresolin, F., Gieren, W., Kudritzki,
R.-P., et al.\ 2009, \apj, 700, 309 
\bibitem[Brooks et al.(2007)]{brooks07} Brooks, A.~M., Governato, F., Booth,
C.~M., et al.\ 2007, \apjl, 655, L17 
\bibitem[Brott \& Hauschildt(2005)] {brott05} Brott, A.~M., Hauschildt, P.~H.,
``A PHOENIX Model Atsmophere Grid for GAIA'', ESAsp, 567, 565
\bibitem[Butler \& Giddings(1985)]{butler85} Butler, K., Giddings, J. 1985,
Newsletter on Analysis of Astronomical Spectra No. 9, University College London
\bibitem[Chiavassa et al.(2011)]{2011A&A...535A..22C} Chiavassa, A., Freytag,
B., Masseron, T., \& Plez, B.\ 2011, \aap, 535, A22 
\bibitem[Colavitti et al.(2008)]{colavitti08} Colavitti, E., Matteucci, F., \&
Murante, G.\ 2008, \aap, 483, 401 
\bibitem[Davidge(2009)]{davidge09} Davidge, T.~J.\ 2009, \apj, 697, 1439 
\bibitem[Dav{\'e} et al.(2011a)]{dave11a} Dav{\'e}, R., Oppenheimer, B.~D., \&
Finlator, K.\ 2011, \mnras, 415, 11 (a) 
\bibitem[Dav{\'e} et al.(2011b)]{dave11b} Dav{\'e}, R., Finlator, K., \&
Oppenheimer, B.~D.\ 2011, \mnras, 416, 1354 (b)
\bibitem[Davies et al.(2009a)]{davies09a} Davies, B., Origlia, L., Kudritzki,
R.~P., et al.\ 2009, \apj, 696, 46 (a)
\bibitem[Davies et al.(2009b)]{davies09b} Davies, B., Origlia, L., Kudritzki,
R.~P., et al.\ 2009, \apj, 696, 2014 (b)
\bibitem[Davies et al.(2010)]{davies10} Davies, B., Kudritzki, R.~P., \& Figer,
D.~F.\ 2010, \mnras, 407, 1203 (b)
\bibitem[De Lucia et al.(2004)]{delucia04} De Lucia, G., Kauffmann, G., \&
White, S.~D.~M.\ 2004, \mnras, 349, 1101 
\bibitem[de Rossi et al.(2007)]{derossi07} de Rossi, M.~E., Tissera, P.~B., \&
Scannapieco, C.\ 2007, \mnras, 374, 323 
\bibitem[Evans et al.(2011)]{evans11} Evans, C.~J., Davies, B., Kudritzki,
R.~P., et al.\ 2011, \aap, 527, 50 
\bibitem[Finlator \& Dav{\'e}(2008)]{finlator08} Finlator, K., \& Dav{\'e}, R.\
2008, \mnras, 385, 2181 
\bibitem[Garnett \& Shields(1987)]{garnett87} Garnett, D.~R., \& Shields, G.~A.\
1987, \apj, 317, 82 
\bibitem[Garnett et al.(1997)]{garnett97} Garnett, D.~R., Shields, G.~A.,
Skillman, E.~D., Sagan, S.~P., \& Dufour, R.~J.\ 1997, \apj, 489, 63 
\bibitem[Garnett(2004)]{garnett04} Garnett, D.~R.\ 2004, In: Cosmochemistry.~The
melting pot of the elements. Cambridge contempuary astrophysics. Cambridge, UK:
Cambridge University Press, p.171 - 216.
\bibitem[Gehren et al.(2001)]{2001A&A...380..645G} Gehren, T., Korn, A.~J., \&
Shi, J.\ 2001, \aap, 380, 645 
\bibitem[Gehren et al.(2001)]{2001A&A...366..981G} Gehren, T., Butler, K.,
Mashonkina, L., Reetz, J., \& Shi, J.\ 2001, \aap, 366, 981 
\bibitem[Grevesse et al.(2007))]{grevesse07} Grevesse, N., Asplund, M,\& Sauval,
A.~J.\ 2007, \ssr, 130, 105
\bibitem[Gustafsson \& Plez(1992)]{1992iesh.conf...86G} Gustafsson, B., \& Plez,
B.\ 1992, Instabilities in Evolved Super- and Hypergiants, 86
\bibitem[Gustafsson et al.(2008)]{gustafsson08} Gustafsson, B., Edvardsson, B.,
Eriksson, K., Jorgensen, U.~G., Nordlund, A, \& Plez, B.\ 2008, \aap, 486, 951
\bibitem[Hauschildt et al.(1997)]{1997ApJ...488..428H} Hauschildt, P.~H., 
Allard, F., Alexander, D.~R., \& Baron, E.\ 1997, \apj, 488, 428 
\bibitem[Heiter \& Eriksson (2006)]{heiter06} Heiter, U. \& Eriksson, B.\
2006,\aap, 452, 2039
\bibitem[Hou et al.(2000)]{hou00} Hou, J.~L., Prantzos, N., \& Boissier, S.\
2000, \aap, 362, 921 
\bibitem[Kewley \& Ellison(2008)]{kewley08} Kewley, L.~J., \& Ellison, S.~L.\
2008, \apj, 681, 1183 
\bibitem[K{\"o}ppen et al.(2007)]{koeppen07} K{\"o}ppen, J., Weidner, C., \&
Kroupa, P.\ 2007, \mnras, 375, 673 
\bibitem[Kudritzki et al.(2008)]{kud08} Kudritzki, R.-P., Urbaneja, M.~A.,
Bresolin, F., et al.\ 2008, \apj, 681, 269
\bibitem[Kudritzki (2010)]{kud10} Kudritzki, R.~P., Astronomical Notes, 331,
459 
\bibitem[Kudritzki et al.(2012)]{kud11} Kudritzki, R.-P., Urbaneja, M.~A.,
Gazak, Z., et al.\ 2012, \apj, in press
\bibitem[Kupka et al.(2000)]{2000BaltA...9..590K} Kupka, F.~G., Ryabchikova,
T.~A., Piskunov, N.~E., Stempels, H.~C., \& Weiss, W.~W.\ 2000, Baltic
Astronomy, 9, 590 
\bibitem[Kupka et al.(1999)]{1999A&AS..138..119K} Kupka, F., Piskunov, N.,
Ryabchikova, T.~A., Stempels, H.~C., \& Weiss, W.~W.\ 1999, \aaps, 138, 119 
\bibitem[Lequeux et al.(1979)]{lequeux79} Lequeux, J., Peimbert, M., Rayo,
J.~F., Serrano, A., \& Torres-Peimbert, S.\ 1979, \aap, 80, 155 
\bibitem[Lee et al. (2006)]{lee06} Lee, H., Skillman, E.D., Cannon, J.M., et
al.\ 2006, \apj, 647, 970
\bibitem[Lind et al.(2011)]{lind11} Lind, K., Asplund, M., Barklem, P.~S., \&
Belyaev, A.~K.\ 2011, \aap, 528, A103 
\bibitem[Lind et al.(2012)]{lind12} Lind, K., Bergemann, M., Asplund, M.\ 2012,
\mnras, submitted
\bibitem[Maiolino et al.(2008)]{maiolino08} Maiolino, R., Nagao, T., Grazian,
A., et al.\ 2008, \aap, 488, 463 
\bibitem[Mashonkina(1996)]{mash96} Mashonkina, L.~J.\ 1996, 
M.A.S.S., Model Atmospheres and Spectrum Synthesis, 108, 140 
\bibitem[Mashonkina et al.(2011)]{2011A&A...528A..87M} Mashonkina, L., Gehren,
T., Shi, J.-R., Korn, A.~J., \& Grupp, F.\ 2011, \aap, 528, A87 
\bibitem[Meynet \& Maeder(2003)]{meynet03} Meynet, G., \& Maeder, A.\ 2003,
\aap, 404, 975 
\bibitem[Naab \& Ostriker(2006)]{naab06} Naab, T., \& Ostriker, J.~P.\ 2006,
\mnras, 366, 899 
\bibitem[O'Brian et al.(1991)]{1991JOSAB...8.1185O} O'Brian, T.~R., Wickliffe,
M.~E., Lawler, J.~E., Whaling, W., \& Brault, J.~W.\ 1991, Journal of the
Optical Society of America B Optical Physics, 8, 1185
\bibitem[Paturel et al.(2003)]{paturel03} Paturel, G., Petit, C., Prugniel, P.,
et al.\ 2003, \aap, 412, 45 
\bibitem[Piskunov et al.(1995)]{1995A&AS..112..525P} Piskunov, N.~E., Kupka, F.,
Ryabchikova, T.~A., Weiss, W.~W., \& Jeffery, C.~S.\ 1995, \aaps, 112, 525
\bibitem[Plez(1998)]{1998A&A...337..495P} Plez, B.\ 1998, \aap, 337, 495 
\bibitem[Plez(2010)]{plez10} Plez, B.\ 2010, ASPC 425, 124
\bibitem[Prantzos \& Boissier(2000)]{prantzos00} Prantzos, N., \& Boissier, S.\
2000, \mnras, 313, 338 
\bibitem[Przybilla et al.(2006)]{przybilla06} Przybilla, N., Butler, K., Becker,
S.~R., \& Kudritzki, R.~P.\ 2006, \aap, 445, 1099
\bibitem[Ralchenko et al. (2012)]{nist} Ralchenko, Yu., Kramida, A.E., Reader,
J., \& NIST ASD Team (2011). NIST Atomic Spectra Database (ver. 4.1.0),
[Online]. Available: http://physics.nist.gov/asd [2012, March 24]. National
Institute of Standards and Technology, Gaithersburg, MD. 
\bibitem[Ryabchikova et al.(1997)]{1997BaltA...6..244R} Ryabchikova, T.~A., 
Piskunov, N.~E., Kupka, F., \& Weiss, W.~W.\ 1997, Baltic Astronomy, 6, 244 
\bibitem[van Regemorter(1962)]{1962ApJ...136..906V} van Regemorter, H.\ 
1962, \apj, 136, 906
\bibitem[Reetz(1999)]{reetz} Reetz, J.\ 1999, PhD thesis, LMU M\"unchen
\bibitem[Rybicki \& Hummer(1991)]{1991A&A...245..171R} Rybicki, G.~B., \&
Hummer, D.~G.\ 1991, \aap, 245, 171 
\bibitem[Rybicki \& Hummer(1992)]{1992A&A...262..209R} Rybicki, G.~B., \&
Hummer, D.~G.\ 1992, \aap, 262, 209 
\bibitem[S{\'a}nchez-Bl{\'a}zquez et al.(2009)]{sanchez09}
S{\'a}nchez-Bl{\'a}zquez, P., Courty, S., Gibson, B.~K., \& Brook, C.~B.\ 2009,
\mnras, 398, 591 
\bibitem[Santiago-Cort{\'e}s et al.(2010)]{santiago10} Santiago-Cort{\'e}s, M.,
Mayya, Y.~D., \& Rosa-Gonz{\'a}lez, D.\ 2010, \mnras, 405, 1293 
\bibitem[Seaton(1962)]{1962amp..conf..375S} Seaton, M.~J.\ 1962, Atomic and 
Molecular Processes, 375
\bibitem[Skillman(1998)]{skillman98} Skillman, E.~D.\ 1998, Stellar astrophysics
for the local group: VIII Canary Islands Winter School 
of Astrophysics, 457 
\bibitem[Steenbock \& Holweger(1984)]{1984A&A...130..319S} Steenbock, W., \&
Holweger, H.\ 1984, \aap, 130, 319 
\bibitem[Tremonti et al.(2004)]{tremonti04} Tremonti, C.~A., Heckman, T.~M.,
Kauffmann, G., et al.\ 2004, \apj, 613, 898 
\bibitem[Wiersma et al.(2009)]{wiersma09} Wiersma, R.~P.~C., Schaye, J., \&
Smith, B.~D.\ 2009, \mnras, 393, 99 
\bibitem[Yin et al.(2009)]{yin09} Yin, J., Hou, J.~L., Prantzos, N., et al.\
2009, \aap, 505, 497 

\end{thebibliography}
